\documentclass[10pt,twocolumn,twoside,journal]{IEEEtran}
\usepackage{multirow}
\usepackage{graphicx}
\usepackage{amsmath}
\usepackage{color}
\usepackage{epsfig}
\usepackage{amsfonts}
\usepackage{amssymb}
\usepackage{amsthm}
\usepackage[usenames,dvipsnames]{pstricks}
\usepackage{epstopdf}
\usepackage{algorithm}
\usepackage{caption}
\usepackage{subcaption}
\usepackage{mathtools}
\usepackage[noend]{algpseudocode}

\begin{document}
\title{{\huge Performance Analysis of OTFS Modulation with Receive 
Antenna Selection}}
\author{Vighnesh S Bhat, G. D. Surabhi$^\dagger$, and A. Chockalingam \\  
Department of Electrical Communication Engineering, Indian Institute of 
Science, Bangalore 560012 \\
$\dagger$ Multimedia Communications Lab, University of Texas at Dallas, 
Richardson, TX 75080-3021}

\maketitle
\begin{abstract}
\maketitle
In this paper, we analyze the performance of orthogonal time frequency space (OTFS) modulation with antenna selection at the receiver, where $n_s$ out of $n_r$ receive antennas with maximum channel Frobenius norms in the delay-Doppler (DD) domain are selected. Single-input multiple-output OTFS (SIMO-OTFS), multiple-input multiple-output OTFS (MIMO-OTFS), and space-time coded OTFS (STC-OTFS) systems with receive antenna selection (RAS) are considered. We consider these systems without and with phase rotation. Our diversity analysis results show that, with no phase rotation, SIMO-OTFS and MIMO-OTFS systems with RAS are rank deficient, and therefore they do not extract the full receive diversity as well as the diversity present in the DD domain. Also, Alamouti coded STC-OTFS system with RAS and no phase rotation extracts the full transmit diversity, but it fails to extract the DD diversity. On the other hand, SIMO-OTFS and STC-OTFS systems with RAS become full-ranked when phase rotation is used, because of which they extract the full spatial as well as the DD diversity present in the system. Also, when phase rotation is used, MIMO-OTFS systems with RAS extract the full DD diversity, but they do not extract the full receive diversity because of rank deficiency. Simulation results are shown to validate the analytically predicted diversity performance.     
\end{abstract}
{

\section{Introduction}
\label{sec1}
Orthogonal time frequency space (OTFS) modulation is a two-dimensional (2D) modulation scheme proposed in the recent literature to tackle the doubly-dispersive nature of mobile radio channels, caused by multipath  propagation environments \cite{jakes},\cite{otfs1},\cite{otfs2}. Conventional multicarrier modulation schemes such as orthogonal frequency division multiplexing (OFDM) embed information symbols in the time-frequency (TF) domain to mitigate inter-symbol interference (ISI) caused by time dispersion. However, the Doppler shifts encountered in high-mobility channels destroy the orthogonality among subcarriers in OFDM. This results in degraded performance of OFDM systems in time-varying channels \cite{Proakis}. OTFS, on the other hand, places the information symbols in delay-Doppler (DD) domain which result in 2D convolution of the information symbols with the channel in the DD domain. OTFS has been found to perform better than OFDM in high-Doppler communication scenarios, such as high-speed trains and vehicle-to-vehicle/vehicle-to-infrastructure communications. Since the signaling in OTFS is done in the DD domain rather than in the TF domain, the interaction of information symbol and rapidly time-varying channel appear as almost time invariant in the DD domain. Also, because of the constant DD channel gain experienced by a OTFS frame, design of equalizers and channel estimation in DD domain is easy. One more advantage of OTFS is that it can be implemented using existing multicarrier modulation schemes, such as OFDM, with additional pre-processing and post-processing modules {\cite{chl_est2}.

Several papers in the recent literature have investigated many key issues in OTFS such as low-complexity signal detection {\cite{ofdm_otfs2}-\cite{lr6}, channel estimation \cite{emb_pil}-\cite{sparse}, peak-to-average power ratio (PAPR) and pulse shapes \cite{papr1}-\cite{puls_otfs}, and multiple access \cite{ma_patent}-\cite{otfs_noma}. In terms of performance analysis, an asymptotic diversity analysis for OTFS has been carried out in \cite{otfs_div}. It established that the asymptotic diversity order achieved in single-input single-output OTFS (SISO-OTFS) is one for ideal biorthogonal waveforms. In other words, OTFS in its basic form does not extract the diversity present in the DD domain. It also explored a phase rotation scheme using transcendental numbers to extract full diversity in the DD domain. It has also reported diversity orders of $n_r$ and $n_rP$ for multiple-input multiple-output OTFS (MIMO-OTFS) without and with phase rotation, respectively, where $n_r$ and $P$ denote the number of receive antennas and the number of resolvable paths in the DD domain, respectively. The analysis in \cite{otfs_eff_div} on the effective diversity of OTFS using rectangular waveforms and a two-path channel has shown that the number of signal pairs that prevent the achievability of full rank is very small for sufficiently large frame sizes. The analysis in \cite{STC_otfs} for space-time coded OTFS (STC-OTFS) with Alamouti code with two transmit antennas has reported diversity orders of $2n_r$ and $2n_rP$ for STC-OTFS without and with phase rotation, respectively. Because of the good diversity slopes in the finite signal-to-noise ratio (SNR) regime even with small frame sizes, STC-OTFS was suggested to be suited for low-latency applications. 

\begin{figure*}[t]
\centering
\includegraphics[width=15cm, height=1.75cm]{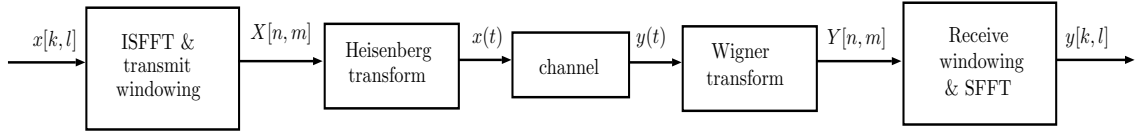}
\caption{OTFS modulation scheme.}
\label{blkdiag}
\end{figure*}

Antenna selection techniques allow the use of fewer radio frequency (RF) chains than the number of antenna elements. This reduces the RF hardware complexity and cost. In this regard, it is of interest to analyze the performance of OTFS with antenna selection, and such an analysis has not been reported so far. Our new and novel contributions in this paper can be highlighted as follows. First, we analyze and establish the diversity orders achieved by different multi-antenna OTFS systems {\em with antenna selection at the receiver}, where $n_s$ out of $n_r$ receive antennas are selected. Second, in rapidly time-varying channels, devising suitable antenna selection metric is a crucial issue. We address this issue by proposing the Frobenius norm of the channel matrix {\em in the DD domain} as the antenna selection criterion. This is novel and attractive because it takes advantage of the simplicity of DD channel estimation in OTFS due to the sparsity and slow variation of rapidly time-varying channels when viewed in the DD domain.

In our analysis, we consider the diversity performance of single-input multiple-output OTFS (SIMO-OTFS), MIMO-OTFS, and STC-OTFS systems with receive antenna selection (RAS). Our diversity analysis results show that, with no phase rotation, SIMO-OTFS and MIMO-OTFS systems with RAS are rank deficient, and therefore they do not extract the full receive diversity as well as the diversity present in the DD domain. Also, Alamouti coded STC-OTFS system with RAS and no phase rotation extracts the full transmit diversity, but it fails to extract the DD diversity. On the other hand, SIMO-OTFS and STC-OTFS systems with RAS become full-ranked when phase rotation is used, because of which they extract the full spatial as well as the DD diversity present in the system. Also, when phase rotation is used, MIMO-OTFS systems with RAS extract the full DD diversity, but they do not extract the full receive diversity because of rank deficiency. A summary of the diversity orders achieved in different multi-antenna OTFS systems with RAS are presented in Table \ref{sumres} in Sec. \ref{sec3}. In the later sections, we will present analytical derivations for the diversity orders in Table \ref{sumres} and supporting simulation results that verify the analytically predicted diversity orders.

The rest of the paper is organized  as follows. The considered multi-antenna OTFS systems with receive antenna selection are presented in Sec. \ref{sec2}. The diversity analyses of these systems for full rank and rank deficient are presented in Sec. \ref{sec3}. Numerical results and discussions are presented 
in Sec. \ref{sec5}. Conclusions are presented in Sec. \ref{sec6}.

\em Notations:} Capital boldface letters denote matrices, lower case boldface letters denote vectors, $\mathbf{\text{diag}}\{x_1,\cdots,x_n\}$ denotes a diagonal matrix with $\{x_1,\cdots,x_n\}$ as its diagonal entries, and $\|\mathbf{X}\|$ denotes the Frobenius norm of matrix $\mathbf{X}$. Transpose and Hermitian operators are denoted by $(\cdot)^T$ and $(\cdot)^H$, respectively.  $|c|$ and $|{\mathbb S}|$ denote the magnitude of the complex scalar $c$ and size of the set ${\mathbb S}$, respectively. ${\mathbb E}[\cdot]$ and $\mbox{Tr}[\cdot]$ denote the expectation and trace operations, respectively. $\mathcal{CN}(a,b)$ denotes complex Gaussian distribution with mean $a$ and variance $b$.

\section{Multi-antenna OTFS systems with RAS}
\label{sec2}
In this section, we present the basic OTFS modulation scheme and the system models corresponding to different multi-antenna OTFS systems. The analyses that follow in Sec. \ref{sec3} are for integer Dopplers/delays, and the case of fractional Doppler/delays will be analyzed in the Appendix.

\subsection{Basic OTFS modulation}
\label{sec2a}
The OTFS modulation scheme consists of cascaded structures of two 2D transforms at the transmitter and the receiver. The block diagram of the basic OTFS modulation scheme is shown in Fig. \ref{blkdiag}. At the transmitter, information symbols in the DD domain are mapped to TF domain using inverse symplectic finite Fourier transform (ISFFT) followed by windowing. The TF symbols are then converted to time domain using Heisenberg transform for transmission over the channel. At the receiver, Wigner transform (inverse of Heisenberg transform) is performed to get TF symbols. Using windowing and symplectic finite Fourier transform (SFFT), TF symbols are mapped back to DD domain for demodulation.

The information symbols $x[k,l]$s are multiplexed on an $N\times M$ DD grid, given by
\begin{equation}
\Gamma=\lbrace (\tfrac{k}{NT},\tfrac{l}{M\Delta f}), k=0,\cdots,
N-1, l=0,\cdots, M-1\rbrace,
\label{DDgrid}
\end{equation}
where  $1/NT$ and $1/M\Delta f$ denote the bin sizes in the Doppler domain and delay domain, respectively, and $N$ and $M$ denote the number of Doppler and delay bins, respectively. The DD domain symbols $x[k,l]$s are mapped to symbols in the TF domain $X[n,m]$s using ISFFT. Assuming rectangular windowing, the TF signal can be written as 
\begin{equation}
X[n,m] = \frac{1}{\sqrt{MN}}\sum_{k=0}^{N-1}\sum_{l=0}^{M-1} x[k,l]e^{j2\pi
\left(\frac{nk}{N}-\frac{ml}{M}\right)}.
\label{otfsmod}
\end{equation}
This TF signal is converted into a time domain signal $x(t)$, using Heisenberg transform and transmit pulse $g_{tx}(t)$, as 
\begin{equation}
x(t)= \sum_{n=0}^{N-1} \sum_{m=0}^{M-1} X[n,m]g_{tx}(t-nT)e^{j2\pi m \Delta f (t-nT)}.
\label{tfmod}
\end{equation}
The transmitted signal $x(t)$ passes through the channel, whose complex baseband channel response in the DD domain, denoted by $h(\tau,\nu)$, is given by \cite{otfs4}
\begin{equation}
h(\tau,\nu) =\sum_{i=1}^{P} h_i \delta(\tau -\tau_{i}) \delta(\nu-\nu_{i}),
\label{sparsechannel}
\end{equation}
where $P$ is the number of paths in the DD domain, and $h_i$, $\tau_i$, and $\nu_i$ denote the channel gain, delay, and Doppler shift, respectively, associated with the $i$th path. The received time domain signal $y(t)$ at the receiver is then given by
\begin{equation}
\label{channel}
y(t)=\int_{\nu} \int_{\tau} h(\tau,\nu)x(t-\tau)e^{j2\pi\nu(t-\tau)} \mathrm{d} \tau \mathrm{d} \nu+v(t),
\end{equation}
where $v(t)$ denotes the additive white Gaussian noise. 

At the receiver, the received signal $y(t)$ is matched filtered with a receive pulse $g_{rx}(t)$, yielding the cross-ambiguity function $A_{g_{rx},y}(t,f)$ given by
\begin{equation}
\label{crossambig}
A_{g_{rx},y}(t,f)=\int g_{rx} ^*(t'-t) y(t') e^{-j2 \pi f(t'-t)} \mathrm{d}t'.
\end{equation}
The pulses $g_{tx}(t)$ and $g_{rx}(t)$ are chosen such that the biorthogonality condition is satisfied, i.e., $A_{g_{rx},g_{tx}}(t,f)|_{nT, m\Delta f}=\delta (m)\delta (n)$. Sampling $A_{g_{rx},y}(t,f)$ at $t=nT$, $f=m \Delta f$ gives
\begin{equation}
\label{wigner}
Y[n,m] = A_{g_{rx},y}(t,f)|_{t=nT,f =m \Delta f}.
\end{equation}
This received TF domain signal $Y[n,m]$ is mapped to the corresponding DD domain signal $y[k,l]$ using SFFT as
\begin{equation}
y[k,l]=\frac{1}{\sqrt{MN}}\sum_{k=0}^{N-1}\sum_{l=0}^{M-1} Y[n,m]e^{-j2\pi\left(\frac{nk}{N}-\frac{ml}{M}\right)}.
\label{otfsdemod}
\end{equation}
From (\ref{tfmod})-(\ref{otfsdemod}), the input-output relation in the DD domain can be written as \cite{otfs4}
\begin{equation}
y[k,l] = \sum_{i=1}^{P} h_i' x[(k-\beta_i)_N,(l-\alpha_i)_M] + v[k,l],
\label{inpopnofracdopp}
\end{equation} 
where $h_i'=h_i e^{-j2 \pi \nu_i \tau_i}$, $\alpha_i$ and $\beta_i$ are assumed to be integers corresponding to the indices of the delay tap and Doppler frequency associated with $\tau_i$ and $\nu_i$, respectively, i.e., $\tau_i\triangleq\frac{\alpha_i}{M\Delta f}$ and $\nu_i\triangleq\frac{\beta_i}{NT}$, $(.)_N$ denotes the modulo $N$ operation, and $v[k,l]$ denotes the additive white Gaussian noise. Vectorizing the input-output relation in (\ref{inpopnofracdopp}),
we can write \cite{otfs4}
\begin{equation}
\mathbf{y} = \mathbf{Hx} + \mathbf{v}, 
\label{vecform}
\end{equation}
where ${\bf H} \in \mathbb{C}^{MN\times MN}$, ${\bf x}, {\bf y},{\bf v} \in \mathbb{C}^{MN\times 1}$, the ($k+Nl$)th entry of ${\bf x}$, $x_{k+Nl}=x[k,l]$, $k=0,\cdots,N-1$, $l=0,\cdots,M-1$ and $x[k,l] \in \mathbb{A}$, where $\mathbb{A}$ is the modulation alphabet (e.g., quadrature amplitude modulation (QAM) or phase shift keying (PSK)). Likewise, $y_{k+Nl}=y[k,l]$ and $v_{k+Nl}=v[k,l]$, $k=0,\cdots,N-1, l=0,\cdots,M-1$. It is assumed that the $h_i$s are i.i.d and are distributed as $\mathcal{CN}(0,1/P)$, assuming uniform scattering profile.

\begin{figure*}[t]
\centering
\includegraphics[width=16.0 cm, height=4.5 cm]{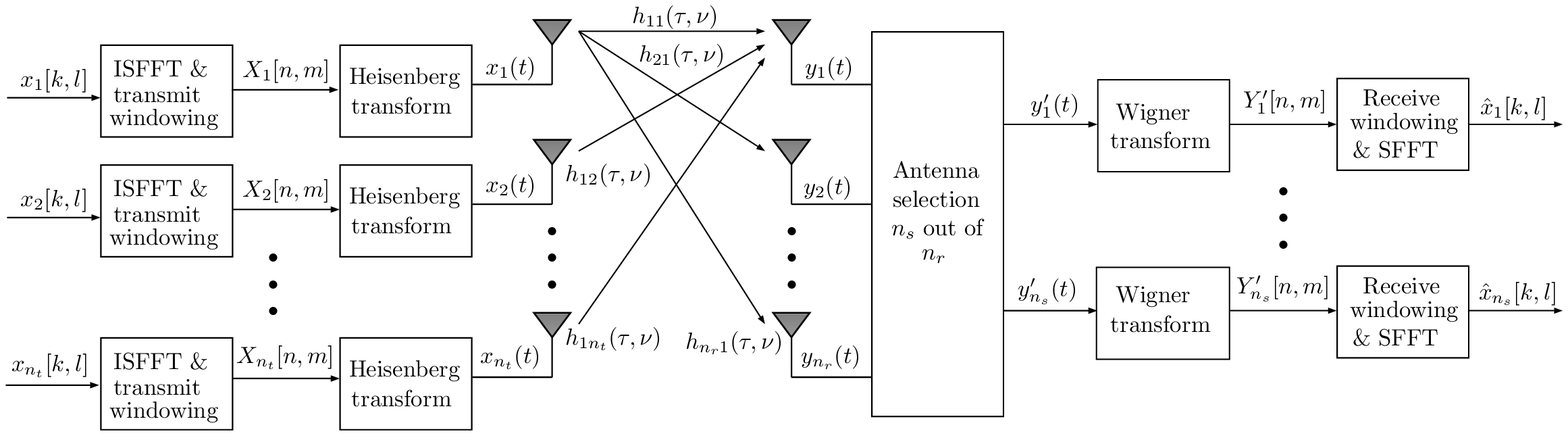}
\caption{MIMO-OTFS with receive antenna selection.}
\label{blkdiag_ant_sel}
\end{figure*}

{\em An alternate form of input-output relation (\ref{vecform}):}
The vectorized form of input-output relation in (\ref{vecform}) can be written in an alternate form which is essential for our diversity analysis. This alternate representation is also useful in writing the system model for STC-OTFS systems. Towards this, it is observed that there are only $P$ non-zero entries in each row and column of the equivalent channel matrix ${\bf H}$ because of the modulo operations in (\ref{inpopnofracdopp}), i.e., there are only $MNP$ non-zero entries in ${\bf H}$. Also, among the non-zero entries there are only $P$ unique values, since each transmitted symbol experiences the same channel gain as can be seen in (\ref{inpopnofracdopp}). With this, the relation in (\ref{vecform}) can be written in an alternate form as \cite{otfs_div}
\begin{equation}
\mathbf{y}^T=\mathbf{h}'\mathbf{X}+\mathbf{v}^T,
\label{hXform}
\end{equation}
where $\mathbf{y}^T$ is $1 \times MN$ received vector, $\mathbf{h}'$ is $1 \times P$ vector whose $i$th entry is given by $h_i'=h_{i}e^{-j2 \pi \nu_i \tau_i}$, $\mathbf{v}^T$ is $1 \times MN$ noise vector, and $\mathbf{X}$ is $P \times MN$ matrix whose $i$th column $\mathbf{X}[i]$, $i=k+Nl, \ k=0,\cdots,N-1, \ l=0,\cdots,M-1$, is given by
\begin{align}
\mathbf{X}[i] & =\begin{bmatrix}
x_{(k-\beta_1)_N+N(l-\alpha_1)_M} \\
x_{(k-\beta_2)_N+N(l-\alpha_2)_M} \\
\vdots \\
x_{(k-\beta_P)_N+N(l-\alpha_P)_M}
\end{bmatrix}.
\label{X_mat}
\end{align}
This representation allows us to view the matrix $\mathbf{X}$ in the form of $P \times MN$ symbol matrix. 

\subsection{MIMO-OTFS with receive antenna selection}
\label{sec2b}
The input-output relation of MIMO-OTFS system with $n_r$  receive antennas and $n_t$ transmit antennas can be written as 
\begin{equation}
\underbrace{
\begin{bmatrix}
\mathbf{y}_1\\ 
\vdots\\
\mathbf{y}_{n_r} 
\end{bmatrix}}_{\triangleq \ \mathbf{\bar{y}}}
=
\underbrace{
\begin{bmatrix}
\mathbf{H}_{11} &\cdots  & \mathbf{H}_{1n_t} \\ 
\vdots & \ddots  & \vdots \\ 
\mathbf{H}_{n_r1}& \cdots  & \mathbf{H}_{n_rn_t}
\end{bmatrix}}_{\triangleq \ \mathbf{\bar{H}}}
\underbrace{
\begin{bmatrix}
\mathbf{x}_1\\ 
\vdots \\ 
\mathbf{x}_{n_t}
\end{bmatrix}}_{\triangleq \ \mathbf{\bar{x}}} 
+
\underbrace{
\begin{bmatrix}
\mathbf{v}_1\\ 
\vdots\\
\mathbf{v}_{n_r} 
\end{bmatrix}}_{\triangleq \ \mathbf{\bar{v}}},
\label{mimo1}
\end{equation}
or equivalently
\begin{equation}
\mathbf{\bar{y}}=\mathbf{\bar{H}}\mathbf{\bar{x}}+\mathbf{\bar{v}},
\label{mimo2}
\end{equation}
where $\mathbf{\bar{y}} \in \mathbb{C}^{{n_r}MN \times 1}$ is the received signal vector, $\mathbf{\bar{H}}  \in \mathbb{C}^{n_{r}MN\times n_{t}MN}$ is the overall equivalent channel matrix with $\mathbf{H}_{ij}$ being the $MN\times MN$ equivalent channel matrix between the $j$th transmit antenna and $i$th receive antenna, $\mathbf{\bar{x}} \in \mathbb{C}^{n_tP \times MN}$ is the OTFS transmit vector, and $\mathbf{\bar{v}}\in \mathbb{C}^{n_rMN \times 1}$ is the noise vector. Perfect DD channel knowledge is assumed at the receiver. The receiver selects $n_s$ out of the $n_r$ antennas with the largest Frobenius norms of the channel in the DD domain, i.e., selects the $n_s$ antennas whose Frobenius norms among those of all the $n_r$ antennas,
given by
\begin{equation}
\sum_{j=1}^{n_t} \|\mathbf{H}_{ij}\|^2, \ i=1,2,\cdots,n_r,
\label{selr1}
\end{equation}
are the largest. Observing that each $\mathbf{H}_{ij}$ contains only $PMN$ non-zero elements with $P$ unique elements and using the definition of Frobenius norm, the selection metric in (\ref{selr1}) can be written as
\begin{equation}
\sum_{k=1}^{P}\sum_{j=1}^{n_t} |h^{(k)}_{ij}|^2, \ i=1,2,\cdots,n_r,
\label{selr2}
\end{equation}
where $h^{(k)}_{ij}$ are the unique non-zero entries of $\mathbf{H}_{ij}$. Therefore, with antenna selection, the input-output relation of the MIMO-OTFS system can be written as 
\begin{equation}
\underbrace{
\begin{bmatrix}
\mathbf{y}'_1\\ 
\vdots\\
\mathbf{y}'_{n_s} 
\end{bmatrix}}_{\triangleq \ \mathbf{\bar{y}}'}
=
\underbrace{
\begin{bmatrix}
\mathbf{H}'_{11} &\cdots  & \mathbf{H}'_{1n_t} \\ 
\vdots & \ddots  & \vdots \\ 
\mathbf{H}'_{n_s1}& \cdots  & \mathbf{H}'_{n_sn_t}
\end{bmatrix}}_{\triangleq \ \mathbf{\bar{H}}'}
\underbrace{
\begin{bmatrix}
\mathbf{x}_1\\ 
\vdots \\ 
\mathbf{x}_{n_t}
\end{bmatrix}}_{\triangleq \ \mathbf{\bar{x}}'}
+
\underbrace{
\begin{bmatrix}
\mathbf{v}'_1\\ 
\vdots\\
\mathbf{v}'_{n_s} 
\end{bmatrix}}_{\triangleq \ \mathbf{\bar{v}}'},
\label{mimosel1}
\end{equation}
or equivalently 
\begin{equation}
\mathbf{\bar{y}'}=\mathbf{\bar{H}'}\mathbf{\bar{x}}+\mathbf{\bar{v}'},
\label{mimosel2}
\end{equation}
where $\mathbf{\bar{y}'} \in \mathbb{C}^{{n_s}MN \times 1}$, $\mathbf{\bar{H}'} \in \mathbb{C}^{n_{s}MN\times n_tMN}$ is the equivalent channel matrix with antenna selection, $\mathbf{\bar{x}} \in \mathbb{C}^{n_tMN \times 1}$ is the OTFS transmit vector, and  $\mathbf{\bar{v}'} \in \mathbb{C}^{{n_s}MN \times 1}$ is the noise vector. Figure \ref{blkdiag_ant_sel} shows the block diagram of MIMO-OTFS with receive antenna selection. 

{\em An alternate form of  MIMO-OTFS with antenna selection:} 
The input-output relation in (\ref{mimosel2}) can be written in an alternate form similar to that in (\ref{hXform}), by observing that each $\mathbf{H}'_{ij}$ in (\ref{mimosel1}) contains only $P$ unique non-zero elements and hence $\mathbf{\bar{H}'}$ in (\ref{mimosel2}) contains only $Pn_sn_t$ unique non-zero elements with each row having only $Pn_t$ unique non-zero elements and each column having only $n_sP$ unique non-zero elements. Therefore, (\ref{mimosel1}) can be written as
\begin{equation}
\underbrace{
\begin{bmatrix}
{\mathbf{y}'_1}^{T}\\ 
\vdots \\ 
{\mathbf{y}'_{n_s}}^{\hspace{-2mm} T}
\end{bmatrix}}_{\triangleq \ \mathbf{\tilde{Y}}}
=
\underbrace{
\begin{bmatrix}
\mathbf{h}'_{11} & \cdots & \mathbf{h}'_{1n_t}  \\ 
\vdots & \ddots  & \vdots \\ 
\mathbf{h}'_{n_s1} & \cdots  & \mathbf{h}'_{n_sn_t}\
\end{bmatrix}}_{\triangleq \ \mathbf{\tilde{H}}}
\underbrace{
\begin{bmatrix}
\mathbf{X}_1\\ 
\vdots \\ 
\mathbf{X}_{n_t}
\end{bmatrix}}_{\triangleq \ \mathbf{\tilde{X}}}
+
\underbrace{
\begin{bmatrix}
{\mathbf{v}'_1}^{\hspace{-0.2mm} T}\\ 
\vdots \\ 
{\mathbf{v}'_{n_s}}^{\hspace{-2mm} T}
\end{bmatrix}}_{\triangleq \ \mathbf{\tilde{V}}},
\label{MIMOmat1}
\end{equation}
or equivalently 
\begin{equation}
\mathbf{\tilde{Y}}=\mathbf{\tilde{H}}\mathbf{\tilde{X}}+\mathbf{\tilde{V}},
\label{MIMOmat2}
\end{equation}
where $\mathbf{\tilde{Y}}\in \mathbb{C}^{n_s \times MN}$ with its $i$th row corresponding to the received signal in the $i$th selected receive antenna, $\mathbf{\tilde{H}} \in \mathbb{C}^{n_s \times n_tP}$ is the channel matrix with $\mathbf{h}'_{ij} \in \mathbb{C}^{1 \times P}$ containing $P$ unique non-zero entries of $\mathbf{H}'_{ij}$, $\mathbf{\tilde{X}}$ is $n_tP \times MN$ symbol matrix, and $\mathbf{\tilde{V}} \in \mathbb{C}^{n_s \times MN}$ is the noise matrix.

\subsection{STC-OTFS with antenna selection}
\label{sec2c}
Figure \ref{stc_otfs} shows the block diagram of STC-OTFS with receive antenna selection. In this subsection, we develop the system model for Alamouti code \cite{alamouti_code} based STC-OTFS with receive antenna selection. 

\subsubsection{Alamouti STC-OTFS}
\label{sec2c1}
Alamouti code based STC-OTFS uses the structure of the well known Alamouti code, generalized to matrices. An STC-OTFS codeword matrix $\tilde{\mathbf{X}}$ is an $n_tMN \times T'MN$ block matrix. Each block in this matrix is an $MN\times MN$ OTFS transmit matrix; e.g., the block $\tilde{\mathbf{X}}_{kt}$ in $\tilde{\mathbf{X}}$ denotes the OTFS transmit matrix in the $t$th frame from $k$th transmit antenna. If $\tilde{\mathbf{X}}$ contains $Z$ independent OTFS symbol matrices which are transmitted over $T'$ frame uses, then the code rate is $Z/T'$ symbols per channel use. A delay-Doppler channel which is quasi-static over $T'$ frame duration is assumed. A $2MN \times 2MN$ Alamouti STC-OTFS codeword matrix with $n_t=T'=2$ is given by \cite{STC_otfs}
\begin{equation}
\tilde{\mathbf{X}}=
\begin{bmatrix}
\mathbf{X}_{1} & -\mathbf{X}_{2}^{H}\\ \\
\mathbf{X}_{2} &  \mathbf{X}_{1}^{H}\\
\end{bmatrix},
\label{hX_stc2}
\end{equation}
where $\mathbf{X}_1$ and $\mathbf{X}_2$ are the symbol matrices. That is, the OTFS transmit vectors corresponding to $\mathbf{X}_1$ and $\mathbf{X}_2$ are transmitted from the 1st and 2nd antennas, respectively, in the first frame. In the second frame, the vectors corresponding to $-\mathbf{X}_2^H$ and $\mathbf{X}_1^H$ are transmitted from the 1st and 2nd antennas, respectively. Following the development of the system model without receive antenna selection in \cite{STC_otfs}, the input-output relation for Alamouti STC-OTFS with selection of $n_s$ out of $n_r$ antennas at the receiver can be written in the form 
\begin{equation}
\begin{split}
\underbrace{
\begin{bmatrix}
\mathbf{y}'_{11}\\ 
\vdots \\ 
\mathbf{y}'_{n_s1}\\ 
(\mathbf{\hat{y}'}_{12})^*\\ 
\vdots \\ 
(\mathbf{\hat{y}'}_{n_s2})^*
\end{bmatrix}}_{\triangleq \ \mathbf{\bar{y}}'} 
=
\underbrace{
\begin{bmatrix} \mathbf{H}'_{11} & \mathbf{H}'_{12}\\ \vdots  &  \vdots \\ \mathbf{H}'_{n_s1} & \mathbf{H}'_{n_s2} \\ \mathbf{H}'^H_{12} & - \mathbf{H'}^H_{11}\\ \vdots & \vdots \\ \mathbf{H'}^H_{n_s2} & -\mathbf{H'}^H_{n_s1} 
\end{bmatrix}}_{\triangleq \ \mathbf{\bar{H}}'}
\underbrace{
\begin{bmatrix}
\mathbf{x}_1\\ 
\mathbf{x}_2
\end{bmatrix}}_{\triangleq \ \mathbf{\bar{x}}}
+
\underbrace{
\begin{bmatrix}
\mathbf{v}'_{11}\\ 
\vdots \\ 
\mathbf{v}'_{n_s1}\\ 
(\mathbf{\hat{v}'}_{12})^*\\ 
\vdots \\ 
(\mathbf{\hat{v}'}_{n_s2})^*
\end{bmatrix}}_{\triangleq \ \mathbf{\bar{v}}'},
\label{STCOTFS}
\end{split}
\end{equation}
where $\mathbf{y}'_{ij} \in \mathbb{C}^{MN \times 1}$ is the received signal vector at the $i$th antenna in the $j$th time slot with $\mathbf{\hat{y}}'_{ij}=\mathbf{Py}'_{ij}$, where $\mathbf{P}$ is $MN \times MN$ permutation matrix given by
\begin{equation}
\mathbf{P}=\mathbf{P}'_M\otimes \mathbf{P}'_N,
\end{equation}
where $\otimes$ denotes the Kronecker product, and $\mathbf{P}'_M$ and $\mathbf{P}'_N$ are left circulant matrices, which are given by
\begin{equation}
\mathbf{P}'_M=\begin{bmatrix}
1 & 0 & \cdots 0 & 0 \\
0 & 0 & \cdots 0 & 1 \\
0 & 0 & \cdots 1 & 0 \\
\vdots \\
0 & 1 & \cdots 0 & 0
\end{bmatrix}_{M\times M}
\hspace{-4.0mm}
\mathbf{P}'_N=\begin{bmatrix}
1 & 0 & \cdots 0 & 0 \\
0 & 0 & \cdots 0 & 1 \\
0 & 0 & \cdots 1 & 0 \\
\vdots \\
0 & 1 & \cdots 0 & 0
\end{bmatrix}_{N\times N}\hspace{-1mm},
\end{equation}
$\mathbf{H'}_{ij} \in \mathbb{C}^{MN \times MN}$ is the equivalent channel matrix between $i$th selected receive antenna and $j$th transmit antenna, and $\mathbf{x}_i \in \mathbb{C}^{MN \times 1}$ is the transmitted OTFS vector. The compact form of (\ref{STCOTFS}) is given by 
\begin{equation}
\mathbf{\bar{y}'}=\mathbf{\bar{H}'}\mathbf{\bar{x}}+\mathbf{\bar{v}'},
\label{STCOTFS1}
\end{equation}
where $\mathbf{\bar{y}'}, \mathbf{\bar{v}'} \in \mathbb{C}^{2n_sMN \times 1}$, $\mathbf{\bar{H}'} \in \mathbb{C}^{2n_sMN \times 2MN}$, and $\mathbf{\bar{x}} \in \mathbb{C}^{2MN \times 1}$.

\begin{figure}[t]
\centering
\includegraphics[width=8.65cm, height=2cm]{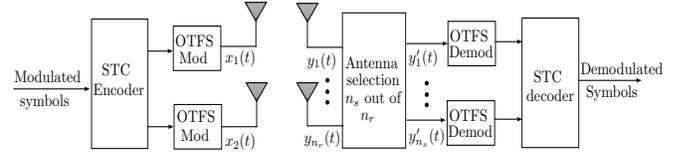}
\caption{STC-OTFS system receive antenna selection.}
\label{stc_otfs}
\vspace{-2mm}
\end{figure}

{\em An alternate form of Alamouti STC-OTFS with antenna selection:}
The input-output relation in (\ref{STCOTFS1}) can be written in an alternate form, based on (\ref{hXform}) and (\ref{MIMOmat1}), as 
\begin{equation}
\begin{split}
\underbrace{
\begin{bmatrix}
\mathbf{y}'^T_{11}  & \mathbf{y}'^T_{12}\\ 
\vdots & \vdots  \\ 
\mathbf{y}'^T_{n_s1} & \mathbf{y}'^T_{n_s2}
\end{bmatrix}}_{\triangleq \ \mathbf{\tilde{Y}}} 
&=
\underbrace{
\begin{bmatrix}
\mathbf{h}'_{11} & \mathbf{h}'_{12}\\ 
\vdots & \vdots \\ 
\mathbf{h}'_{n_s1} & \mathbf{h}'_{n_s2}
\end{bmatrix}}_{\triangleq \ \mathbf{\tilde{H}}}
\underbrace{
\begin{bmatrix}
\mathbf{X}_1 & -(\hat{\mathbf{X}}_2)^*\\ 
{\mathbf{X}}_2 & (\hat{\mathbf{X}}_1)^*
\end{bmatrix}}_{\triangleq \ \mathbf{\tilde{X}}}
\\ & \quad \ + 
\underbrace{ 
\begin{bmatrix}
\mathbf{v}'^T_{11}  & \mathbf{v}'^T_{12}\\ 
\vdots & \vdots  \\ 
\mathbf{v}'^T_{n_s1} & \mathbf{{v}'}^T_{n_s2}
\end{bmatrix}}_{\triangleq \ \mathbf{\tilde{V}}},
\label{STCOTFSMF1}
\end{split}
\end{equation}
which can be written in the following compact form\footnote{In order to adopt a unified input-output system model in the analysis, we keep the same notation in (\ref{MIMOmat2}) and (\ref{STCOTFSMF2}), where in MIMO-OTFS without space-time coding in (\ref{MIMOmat2}), we have $\mathbf{\tilde{Y}}$, $\mathbf{\tilde{V}}$  $\in \mathbb{C}^{n_s \times MN}$, $\mathbf{\tilde{H}} \in \mathbb{C}^{n_s \times n_tP} $ and $\mathbf{\tilde{X}} \in \mathbb{C}^{n_tP \times MN}$, and in space-time coded OTFS in (\ref{STCOTFSMF2}), we have $\mathbf{\tilde{Y}}$, $\mathbf{\tilde{V}}$ $\in \mathbb{C}^{n_s \times 2MN}$, $\mathbf{\tilde{H}} \in \mathbb{C}^{n_s \times 2P} $ and $\mathbf{\tilde{X}} \in \mathbb{C}^{2P \times 2MN} $.}: 
\begin{equation}
\mathbf{\tilde{Y}}=\mathbf{\tilde{H}}\mathbf{\tilde{X}}+\mathbf{\tilde{V}},
\label{STCOTFSMF2}
\end{equation}
where $\mathbf{\tilde{Y}}, \mathbf{\tilde{V}} \in \mathbb{C}^{n_s \times 2MN}$, $\mathbf{\tilde{H}} \in \mathbb{C}^{n_s \times 2P}$, and $\mathbf{\tilde{X}} \in \mathbb{C}^{2P \times 2MN}$. Here it is observed that $\mathbf{X}_i \neq \mathbf{\hat{X}}_i$, since transmitted OTFS vectors in the 2nd frame are conjugated and permuted vectors of those transmitted in the 1st frame. \textcolor{black}{In (\ref{hX_stc2})}, $\mathbf{\tilde{X}}$ is defined to be \textcolor{black}{$2MN \times 2MN$} symbol matrix, but for diversity analysis $\mathbf{\tilde{X}} \in \mathbb{C}^{2P \times 2MN}$ in (\ref{STCOTFSMF1}) is convenient.

\subsection{OTFS with phase rotation}
\label{sec2d}
In this subsection, we present OTFS modulation with phase rotation. In OTFS with phase rotation, the OTFS vector ${\bf x}$ is pre-multiplied by a phase rotation matrix $\mathbf{\Phi}$, which is of the form
\begin{equation}
\mathbf{\Phi }=\text{diag}\{\phi _0,\phi _1,\cdots ,\phi_{MN-1}\}.
\label{theorem1}
\end{equation}
That is, $\mathbf{x}'=\mathbf{\Phi }\mathbf{x}$ is the phase rotated OTFS transmit vector. It has been shown in \cite{otfs_div} that SISO-OTFS with the above phase rotation achieves the full diversity available in the DD domain when $\phi_i=e^{ja_i}$, $i=0,1,\cdots,MN-1$, are transcendental numbers with $a_i$ being real, distinct and algebraic. We consider this phase rotation scheme for multi-antenna OTFS systems, where the OTFS vector in each transmit antenna is pre-multiplied by the phase rotation matrix $\mathbf{\Phi}$. 

\subsection{Rank of multi-antenna OTFS systems}
\label{sec2e}
In the next section, we carry out the diversity analysis for multi-antenna systems for full rank and rank deficient cases. In this subsection, we identify the rank of the considered multi-antenna OTFS systems without and with phase rotation.

\subsubsection{MIMO-OTFS, SIMO-OTFS}
\label{sec2e1}
Consider MIMO-OTFS ($n_t\geq 2$) without phase rotation. Let $\mathbf{\tilde{X}}_i$ and $\mathbf{\tilde{X}}_j$ be two distinct symbol matrices defined in (\ref{MIMOmat2}). The minimum rank of $(\mathbf{\tilde{X}}_i-\mathbf{\tilde{X}}_j)$ is $1 < \min(n_tP,MN)$ \cite{otfs_div}. Therefore, MIMO-OTFS without phase rotation is rank deficient. Next, consider MIMO-OTFS with phase rotation. Let $\mathbf{\bar{x}}'_i=\Phi \mathbf{\bar{x}}_i$ and $\mathbf{\bar{x}}'_j=\Phi \mathbf{\bar{x}}_j$ be two distinct phase rotated OTFS transmit vectors in (\ref{mimosel2}). Let $\mathbf{\tilde{X}}'_i$ and $\mathbf{\tilde{X}}'_j$ be the corresponding phase rotated symbol matrices in (\ref{MIMOmat2}). The minimum rank of $(\mathbf{\tilde{X}}'_i-\mathbf{\tilde{X}}'_j)$ is $P < \min(n_tP,MN)$ \cite{otfs_div}. Therefore, MIMO-OTFS system with phase rotation is also rank deficient.

SIMO-OTFS can be viewed as a special case of MIMO-OTFS with $n_t=1$. Therefore, for SIMO-OTFS without phase rotation, the minimum rank of $(\mathbf{\tilde{X}}_i-\mathbf{\tilde{X}}_j)$ is $1<\min(P,MN)$. Therefore, SIMO-OTFS system without phase rotation is rank deficient for $P > 1$. For $P=1$, the dimension of $(\mathbf{\tilde{X}}_i-\mathbf{\tilde{X}}_j)$ is $1 \times MN$ and the minimum rank is $1$, and so it is full rank. For SIMO-OTFS with phase rotation, the minimum rank of $(\mathbf{\tilde{X}}'_i-\mathbf{\tilde{X}}'_j)$ is $P =\min(P,MN)$. Since $P \leq MN$ and minimum rank is $P$, and so it is full rank.

\subsubsection{STC-OTFS}
\label{sec2e3}
Consider Alamouti STC-OTFS without phase rotation. Let 
$\mathbf{\tilde{X}}_i$ and $\mathbf{\tilde{X}}_j$ be the two distinct symbol matrices defined in (\ref{STCOTFSMF2}), The minimum rank of $(\mathbf{\tilde{X}}_i-\mathbf{\tilde{X}}_j)$ is $2 < \min(2P,2MN)$ \cite{STC_otfs}. Therefore, for $P>1$ Alamouti STC-OTFS is rank deficient, and for $P=1$ it is full rank with rank $2$. For Alamouti STC-OTFS with phase rotation, the minimum rank of $(\mathbf{\tilde{X}}'_i-\mathbf{\tilde{X}}'_j)$ is $2P \leq \min(2P,2MN)$ \cite{STC_otfs}. Therefore, Alamouti STC-OTFS with phase rotation is full rank with rank $2P$.

\section{Analysis of Multi-antenna OTFS with RAS}
\label{sec3}
In this section, we analyze the performance of multi-antenna OTFS systems with RAS by deriving explicit upper bounds on pairwise error probability (PEP). We carry out the diversity analysis for full rank and rank deficient cases in the following subsections.

\subsection{Full rank multi-antenna OTFS systems with RAS}
\label{sec3a}
Consider the case of full rank multi-antenna OTFS systems with receive antenna selection. Let $\mathbf{\tilde{X}}_i$ and $\mathbf{\tilde{X}}_j$ be two distinct symbol matrices. Assuming perfect DD channel knowledge and maximum likelihood (ML) detection at the receiver, the conditional PEP between the symbol matrices $\mathbf{\tilde{X}}_i$ and $\mathbf{\tilde{X}}_j$, assuming $\mathbf{\tilde{X}}_i$ to be the transmitted symbol matrix, is given by
\begin{equation} 
P(\mathbf {\tilde{X}}_i \rightarrow \mathbf {\tilde{X}}_j|\mathbf {\tilde{H}},\mathbf {\tilde{X}}_i)=Q \left ({\sqrt {\frac {\|\mathbf {\tilde{H}}(\mathbf {\tilde{X}}_i-\mathbf {\tilde{X}}_j)\|^{2}}{2N_0}} }\right)\!,
\label{PEPM0}
\end{equation} 
where $Q(x)=\frac{1}{\sqrt{2\pi}}\int_{x}^{\infty}e^{-t^2/2} dt$.
For convenience, the entries of $\mathbf{\tilde{X}}$ are normalized so that average energy per symbol time is one and the SNR, denoted by $\gamma$, is given by $\gamma=1/N_{0}$. Therefore, (\ref{PEPM0}) can be written as
\begin{equation} 
P(\mathbf {\tilde{X}}_i \rightarrow \mathbf {\tilde{X}}_j|\mathbf {\tilde{H}},\mathbf {\tilde{X}}_i)=Q \left ({\sqrt {\frac {\gamma \|\mathbf {\tilde{H}}(\mathbf {\tilde{X}}_i-\mathbf {\tilde{X}}_j)\|^{2}}{2}} }\right)\!.
 \label{PEPM1}
\end{equation}  
Averaging over the distribution of $\mathbf{\mathbf{\tilde{H}}}$ and upper bounding using Chernoff bound, an upper bound on the unconditional PEP can be written as 
\begin{equation} 
P(\mathbf {\tilde{X}}_i\rightarrow \mathbf {\tilde{X}}_j) \leq \mathbb {E}_{\mathbf{\tilde{H}}} \left [{ {\exp} \left (-{{\frac {\gamma ~\|\mathbf {\tilde{H}} (\mathbf {\tilde{X}}_i-\mathbf {\tilde{X}}_j)\|^{2}}{4}}\ }\right) }\right]\!.
\label{PEPM2}
\end{equation} 
The distribution of $\mathbf{\tilde{H}}$ is given by \cite{MIMO_div},\cite{errata}
\begin{equation} 
\begin{multlined}
f_{\mathbf{\tilde{H}}}(\mathbf{h}'_1,\cdots,\mathbf{h}'_{n_s}) = \frac{n_{r}!}{(n_r-n_s)!n_s!} \\ \hspace{-5mm} \cdot \biggl ( \sum_{l=1}^{n_s} \left[1-e^{-P\|\mathbf{h}'_l\|^{2} } \sum_{k=0}^{n_tP-1}\frac{P^k\|\mathbf{h}'_l\|^{2k}}{k!} \right]^{n_r-n_s} \\ \hspace{-0mm} \cdot \textit{I}_{\mathcal{\tilde{H}}_l}(\mathbf{h}'_1,\cdots,\mathbf{h}'_{n_s}) \biggr )   \cdot \frac{P^{Pn_tn_s}}{\pi^{Pn_tn_s}}e^{-P(\|\mathbf{h}'_1\|^{2}+\cdots+\|\mathbf{h}'_{n_s}\|^{2})},
\label{distm}
\end{multlined}
\end{equation}
where $\mathbf{h}'_i$ is the $i$th row of $\mathbf{\tilde{H}}$, $\textit{I}_{\mathcal{\tilde{H}}_l}(\mathbf{h}'_1,\cdots,\mathbf{h}'_{n_s})$ is the indicator function given by
\begin{equation} 
\textit{I}_{\mathcal{\tilde{H}}_l}(\mathbf{h}'_1,\cdots,\mathbf{h}'_{n_s})=\begin{cases}
1 & \text{ if } (\mathbf{h}'_1,\cdots,\mathbf{h}'_{n_s})\in \mathcal{\tilde{H}}_l\\ 
0 & \text{ else}, 
\end{cases}
\label{ind}
\end{equation} 
and the region $\mathcal{\tilde{H}}_l$ is defined as
$\mathcal{\tilde{H}}_l = \{ \mathbf{h}'_1,\cdots,\mathbf{h}'_{n_s}: \|\mathbf{h}'_l\| < \|\mathbf{h}'_k\|, k=1,\cdots,l-1,l+1,\cdots,n_s \}$. The PEP bound can be written as
\begin{equation} 
\begin{split}
P(\mathbf {\tilde{X}}_i \rightarrow \mathbf {\tilde{X}}_j) &\leq \sum_{l=1}^{n_s}\int_{\mathcal{\tilde{H}}_l}^{}e^{\frac{-\gamma}{4}\|\mathbf{\tilde{H}}(\mathbf{\tilde{X}}_i-\mathbf{\tilde{X}}_j)\|^2}  \frac{n_r!}{(n_r-n_s)!n_s!}\\ & \hspace{-10mm} \cdot \left ( 1-e^{-P\|\mathbf{h}'_l\|^2} \sum_{k=0}^{n_tP-1} \frac{P^k\|\mathbf{h}'_l\|^{2k}}{k!}\right )^{n_r-n_s} \\ & \hspace{-10mm} \cdot \frac{P^{n_sn_tP}}{\pi ^{n_sn_tP}}e^{-P(\|\mathbf{h}'_1\|^2+\cdots+\|\mathbf{h}'_{n_s}\|^2)}d\mathbf{h}'_1 \cdots d\mathbf{h}'_{n_s}.
\label{PEPM3}
\end{split}
\end{equation}
Letting $\sqrt{P}\mathbf{h}'_l=\mathbf{s}_l$, $l=1,\cdots,n_s$, ${\bf S}$ to be an $n_s\times n_tP$ matrix whose $l$th row is ${\bf s}_l$ and region $\mathcal{\tilde{H}}_l = \{ \mathbf{s}_1,\cdots,\mathbf{s}_{n_s}: \|\mathbf{s}_l\| < \|\mathbf{s}_k\|, k=1,\cdots,l-1,l+1,\cdots,n_s \}$, we can write (\ref{PEPM3}) as 
\begin{equation} 
\begin{split}
P(\mathbf {\tilde{X}}_i \rightarrow \mathbf {\tilde{X}}_j) &\leq \sum_{l=1}^{n_s}\int_{\mathcal{\tilde{H}}_l}^{}e^{\frac{-\gamma}{4P}\|\mathbf{S}(\mathbf{\tilde{X}}_i-\mathbf{\tilde{X}}_j)\|^2}  \frac{n_r!}{(n_r-n_s)!n_s!} \\ 
& \hspace{-4mm} \cdot \left (1-e^{-\|\mathbf{s}_l\|^2} \sum_{k=0}^{n_tP-1} \frac{\|\mathbf{s}_l\|^{2k}}{k!}\right )^{n_r-n_s} \\ 
& \hspace{-4mm} \cdot \frac{(\sqrt{P})^{n_sn_tP}}{\pi ^{n_sn_tP}}e^{-(\|\mathbf{s}_1\|^2+\cdots+\|\mathbf{s}_{n_s}\|^2)}d\mathbf{s}_1 \cdots d\mathbf{s}_{n_s}.
\label{PEPM4}
\end{split}
\end{equation}
The term $\|\mathbf{S}(\mathbf{\tilde{X}}_i-\mathbf{\tilde{X}}_j)\|^2$ in (\ref{PEPM4}) can be simplified as
\begin{eqnarray} 
\|\mathbf{S}(\mathbf{\tilde{X}}_i-\mathbf{\tilde{X}}_j)\|^2 {} & = & \mbox{Tr}{ \{ \mathbf{S}(\mathbf{\tilde{X}}_i-\mathbf{\tilde{X}}_j)(\mathbf{\tilde{X}}_i-\mathbf{\tilde{X}}_j)^H\mathbf{S}^H \}} \nonumber \\
& = & \mbox{Tr}{ \{\mathbf{SU\Lambda} \mathbf{(SU)}^H \}} \nonumber \\
& = & \sum_{k=1}^{n_tP}\lambda_k \|\mathbf{c}_k\|^2,
\label{PEPM5}
\end{eqnarray} 
where (\ref{PEPM5}) uses the eigenvalue decomposition of $(\mathbf{\tilde{X}}_i-\mathbf{\tilde{X}}_j)(\mathbf{\tilde{X}}_i-\mathbf{\tilde{X}}_j)^H$, $\mathbf{U}$ is the unitary matrix whose columns are the eigenvectors of $(\mathbf{\tilde{X}}_i-\mathbf{\tilde{X}}_j)(\mathbf{\tilde{X}}_i-\mathbf{\tilde{X}}_j)^H$, $\mathbf{\Lambda}$ is the diagonal matrix containing its eigenvalues, and $\mathbf{c}_k$ is the $k$th column of $\mathbf{SU}$. Let $\mathbf{c}_l'$ be the $l$th row of $\mathbf{SU}$ so that $\mathcal{\tilde{H}}_l = \{ \mathbf{c}'_1,\cdots,\mathbf{c}'_{n_s}: \|\mathbf{c}'_l\| < \|\mathbf{c}'_k\|, k=1,\cdots,l-1,l+1,\cdots,n_s \}$. Defining $K \triangleq n_tP$ and $\rho \triangleq (\sqrt{P})^{n_sn_tP}$, and changing variables in (\ref{PEPM4}) by substituting $c_{ij}=s_{ij}$ for $i=1,\cdots,n_s$ and $j=1,\cdots,n_tP$, we get 
\begin{multline}
P(\mathbf {\tilde{X}}_i \rightarrow \mathbf {\tilde{X}}_j) \leq \frac{\rho \cdot n_r!}{(n_r-n_s)!n_s!} \\
\cdot \sum_{l=1}^{n_s} \int_{\mathcal{\tilde{H}}_l}^{} 
e^{\frac{-\gamma}{4P}\left(\lambda _1(|c_{11}|^2+\cdots+|c_{n_s1}|^2)+\cdots+\lambda_{K}(|c_{1K}|^2+\cdots+|c_{n_sK}|^2)\right )}  \\
\cdot \left ( 1-e^{-(|c_{l1}|^2+\cdots+|c_{lK}|^2)} \hspace{-1mm} \sum_{k=0}^{K-1}\frac{(|c_{l1}|^2+\cdots+|c_{lK}|^2)^k}{k!} \right )^{n_r-n_s} \\
\cdot \frac{1}{\pi^{n_sK}} e^{-\sum_{i=1}^{n_s}\sum_{j=1}^{K}|c_{ij}|^2} dc_{11} \cdots dc_{n_sK}.
\label{PEPM6}
\end{multline}
Evaluating the integral in (\ref{PEPM6}) over the region is difficult. But because of symmetry of pdf it is possible to evaluate over the whole space which results in an upper bound. Because of the symmetry of the pdf, the integral over $\mathcal{\tilde{H}}_l$ for each $l$ is same. The $l$th term in (\ref{PEPM6}) can be rewritten using standard integration as 
\vspace{-4mm}
\begin{multline}
\mathcal{I}_l = \frac{\rho \cdot n_r!}{(n_r-n_s)!n_s!} 
\int_{0}^{\infty} \cdots  \int_{0}^{\infty} \\
\cdot e^{\frac{-\gamma}{4P}\left(\lambda _1(|c_{11}|^2+\cdots+|c_{n_s1}|^2)+\cdots+\lambda_{K}(|c_{1K}|^2+\cdots+|c_{n_sK}|^2)\right )}  \\
\cdot \left ( 1-e^{-(|c_{l1}|^2+\cdots+|c_{lK}|^2)} \hspace{-1mm} \sum_{k=0}^{K-1}\frac{(|c_{l1}|^2+\cdots+|c_{lK}|^2)^k}{k!} \right )^{n_r-n_s} \\
\cdot \frac{1}{\pi^{n_sK}} e^{-\sum_{i=1}^{n_s}\sum_{j=1}^{K}|c_{ij}|^2} dc_{11} \cdots dc_{n_sK}.
\label{PEPM7}
\end{multline}
Changing the variables $c_{ij} =\sigma_{ij}e^{j\theta_{ij}}$, $i=1,\cdots,n_s$, $j=1,\cdots,K$ (with differential element $dc_{ij}=\sigma_{ij}d\sigma_{ij}d\theta_{ij}$), after evaluating integral w.r.t $d\theta_{ij}$ over $[0,2\pi]$, we get
\begin{multline}
\mathcal{I}_l = \frac{2^{n_sK}\rho \cdot n_r!}{(n_r-n_s)!n_s!} \\
\cdot \int_{0}^{\infty} \cdots \int_{0}^{\infty}  
e^{\frac{-\gamma}{4P}\left(\lambda _1(\sigma_{11}^2+\cdots+\sigma_{n_s1}^2)+\cdots+\lambda_{K}(\sigma _{1K}^2+\cdots+\sigma_{n_sK}^2) \right)} \\
 \cdot  \left ( 1-e^{-(\sigma_{l1}^2+\cdots+\sigma_{lK}^2)} \sum_{k=0}^{K-1}\frac{(\sigma_{l1}^2+\cdots+\sigma_{lK}^2)^k}{k!} \right )^{n_r-n_s}\\  
\cdot  e^{-\sum_{i=1}^{n_s}\sum_{j=1}^{K}\sigma_{ij}^2}\sigma_{11} \cdots \sigma_{n_sK} d\sigma_{11} \cdots d\sigma_{n_sK}.
\label{PEPM8}
\end{multline}
Substituting $\sigma^2_{ij}=v_{ij}$, $i=1,\cdots,n_s$, $j=1,\cdots,K$, we get
\begin{multline}
\mathcal{I}_l = \frac{\rho \cdot n_r!}{(n_r-n_s)!n_s!} \\
\cdot \int_{0}^{\infty} \cdots \int_{0}^{\infty}
 e^{\frac{-\gamma}{4P}\left(\lambda _1(v_{11}+\cdots+v_{n_s1})+\cdots+\lambda_{K}(v_{1K}+\cdots+v_{n_sK}) \right)} \\ 
\cdot  \left ( 1-e^{-(v_{l1}+\cdots+v_{lK})} \sum_{k=0}^{K-1}\frac{(v_{l1}+\cdots+v_{lK})^k}{k!} \right )^{n_r-n_s} \\ 
\cdot   e^{-\sum_{i=1}^{n_s}\sum_{j=1}^{K}v_{ij}} dv_{11} \cdots dv_{n_sK}.
\label{PEPM9}
\end{multline}
Now, (\ref{PEPM9}) can be written as
\begin{equation} 
\begin{split}
\mathcal{I}_l &=\frac{\rho \cdot n_r!}{(n_r-n_s)!n_s!}\int_{0}^{\infty} \cdots \int_{0}^{\infty} \\ &  e^{\frac{-\gamma}{4P}\sum_{i=1}^{K}\lambda _i\left(\sum_{d=1,d\neq l}^{n_s}v_{di}\right)}\cdot e^{-\left (\sum_{i=1}^{K}\sum_{d=1,d \neq l}^{n_s} v_{di} \right )}  \\ 
& \cdot \prod_{i=1}^{K}\prod_{d=1,d\neq l}^{n_s}dv_{di} \cdot \int_{0}^{\infty}\cdots \int_{0}^{\infty} e^{\frac{-\gamma}{4P}\sum_{i=1}^{K}\lambda_i v_{li}} \\ 
& \cdot \left (1-e^{-(v_{l1}+\cdots+v_{lK})} \sum_{k=0}^{K-1}\frac{\left ( v_{l1}+\cdots+v_{lK} \right )^k}{k!}\right )^{n_r-n_s} \\ 
& \cdot e^{-(v_{l1}+\cdots+v_{lK})} dv_{l1} \cdots dv_{lK}.
\label{PEPM10}
\end{split}
\end{equation}

\vspace{-3mm}
\hspace{-4mm}
Let $\mathcal{I}_l^{(1)}$ denote the first integral and $\mathcal{I}_l^{(2)}$ denote the second integral in the above expression. Evaluating $\mathcal{I}_l^{(1)}$ using $\int_{0}^{\infty}e^{-\alpha x}dx=\frac{1}{\alpha }$, we get 
\begin{equation} 
\begin{split}
\mathcal{I}_l^{(1)}=\left ( \frac{1}{\prod_{i=1}^{K}(1+\frac{\gamma \lambda_i}{4P})} \right )^{n_s-1}, 
\label{PEPM11}
\end{split}
\end{equation} 
and $\mathcal{I}_l^{(2)}$ as
\begin{equation} 
\begin{split}
\mathcal{I}_l^{(2)} &=\int_{0}^{\infty}\cdots \int_{0}^{\infty} e^{\frac{-\gamma}{4P}\sum_{i=1}^{K}\lambda_i v_{li}} \\ &  \cdot \left ( 1-e^{-(v_{l1}+\cdots+v_{lK})}  \sum_{k=0}^{K-1}\frac{\left ( v_{l1}+\cdots+v_{lK} \right )^k}{k!}\right )^{n_r-n_s} \\ &  \cdot e^{-(v_{l1}+\cdots+v_{lK})} dv_{l1} \cdots dv_{lK}.
\label{PEPM12}
\end{split}
\end{equation}
Let $g(u)=1-e^{-u}\sum_{m=0}^{K-1}\frac{u^m}{m!}$ be the incomplete Gamma function satisfying $g(u)\leq \frac{u^K}{K!}$ for $u>0$. Upper bounding the RHS of (\ref{PEPM12}) by $\frac{u^K}{K!}$ with $u=v_{l1} +\cdots  +v_{lK}$ in (\ref{PEPM12}), we can write  
\begin{equation} 
\begin{split}
\mathcal{I}_l^{(2)} &\leq \frac{1}{{(K!)^{n_r-n_s}}}\int_{0}^{\infty}\cdots \int_{0}^{\infty} e^{\frac{-\gamma}{4P}\sum_{i=1}^{K}\lambda_i v_{li}} \\ & \cdot (v_{l1}+\cdots+v_{lK})^{K(n_r-n_s)} e^{-(v_{l1}+\cdots+v_{lK})} dv_{l1} \cdots dv_{lK}.
\label{PEPM13}
\end{split}
\end{equation}
We observe that 
\begin{eqnarray}
(v_{l1}+\cdots+ v_{lK})^{K(n_r-n_s)} & = & \nonumber \\
&\hspace{-30mm} & \hspace{-30mm} \sum_{i_1=1}^{K}\cdots \sum_{i_{K(n_r-n_s)}=1}^{K} \hspace{-2mm} v_{li_1} \cdots v_{li_{K(n_r-n_s)}},
\label{grp1}
\end{eqnarray}
where index $i_k$ in $v_{li_{k}}$ takes values from the set $\zeta =\{1,\cdots,K \}$ with $k  \in \{1,\cdots,K(n_r-n_s)\}$. Let the index $j$ appear $m_j$ times among the subscripts of the term $v_{li_1} \cdots v_{li_{K(n_r-n_s)}}$ in (\ref{grp1}). Then, 
\begin{equation} 
\begin{split}
v_{li_1} \cdots  v_{li_{K(n_r-n_s)}}  = \prod_{j=1}^{K} (v_{lj})^{m_j}
\end{split}
\label{grp2}
\end{equation}
such that $\sum_{j=1}^{K}m_j=K(n_r-n_s)$. Using (\ref{grp1}) and (\ref{grp2}) in (\ref{PEPM13}) and changing the order of summation and integration, we get
\begin{equation} 
\begin{split}
\mathcal{I}_l^{(2)} &\leq \frac{1}{{(K!)^{n_r-n_s}}} \sum_{i_1=1}^{K}\cdots \sum_{i_{K(n_r-n_s)}=1}^{K} \\ & \hspace{-8mm}\cdot \biggl(   \int_{0}^{\infty}\cdots \int_{0}^{\infty} e^{-\sum_{i=1}^{K}(\frac{\gamma \lambda_i}{4P}+1)v_{li}} \prod_{i=1}^{K}(v_{li})^{m_i} dv_{l1} \cdots dv_{lK} \biggr).
\label{PEPM14}
\end{split}
\end{equation}
Using $\int_{0}^{\infty} x^n e^{-ax} dx=\frac{n!}{a^{n+1}}$, (\ref{PEPM14}) 
can be written as 
\begin{eqnarray} 
\mathcal{I}_l^{(2)} & \leq & \frac{1}{{(K!)^{n_r-n_s}}}\biggl( \sum_{i_1=1}^{K}\cdots \sum_{i_{K(n_r-n_s)}=1}^{K} \nonumber \\  
& & \cdot \frac{m_1!\cdots m_K!}{(1+\frac{\gamma \lambda_1}{4P})^{m_1+1} \cdots (1+\frac{\gamma \lambda_K}{4P})^{m_K+1} }\biggr).
\label{PEPM15}
\end{eqnarray}
Using (\ref{PEPM15}) and (\ref{PEPM11}) in (\ref{PEPM10}), $\mathcal{I}_{l}$ can be written as
\begin{equation} 
\begin{split}
\mathcal{I}_{l} &\leq \frac{\rho \cdot n_r!}{(n_r-n_s)!n_s! (K!)^{n_r-n_s}}  \left ( \frac{1}{\prod_{i=1}^{K}(1+\frac{\gamma \lambda_i}{4P})} \right )^{n_s-1} \\ &  \sum_{i_1=1}^{K}\cdots \sum_{i_{K(n_r-n_s)}=1}^{K}  \frac{m_1!\cdots m_K!}{(1+\frac{\gamma \lambda_1}{4P})^{m_1+1} \cdots (1+\frac{\gamma \lambda_K}{4P})^{m_K+1}}.
\label{PEPM16}
\end{split}
\end{equation}
The above bound is independent of $l$. Therefore, substituting (\ref{PEPM16}) in (\ref{PEPM6}), we can write
\begin{multline}
P(\mathbf {\tilde{X}}_i \rightarrow \mathbf {\tilde{X}}_j)  \\ \leq \frac{\rho \cdot n_r!}{(n_r-n_s)!(n_s-1)! (K!)^{n_r-n_s}}  \left ( \frac{1}{\prod_{i=1}^{K}(1+\frac{\gamma \lambda_i}{4P})} \right )^{n_s-1} \\  \cdot \sum_{i_1=1}^{K}\cdots \sum_{i_{K(n_r-n_s)}=1}^{K}  \frac{m_1!\cdots m_K!}{(1+\frac{\gamma \lambda_1}{4P})^{m_1+1} \cdots (1+\frac{\gamma \lambda_K}{4P})^{m_K+1}}.
\label{PEPM17}
\end{multline}
In the high SNR regime, with some algebraic manipulations, we can write 
\begin{equation} 
\begin{split}
P(\mathbf {\tilde{X}}_i \rightarrow \mathbf {\tilde{X}}_j)  &\leq \frac{\rho \cdot n_r!}{(n_r-n_s)!(n_s-1)! (K!)^{n_r-n_s}}   \frac{1}{\left(\prod_{i=1}^{K}\lambda_i \right)^{n_s}} \\ & \cdot \biggl(  \sum_{i_1=1}^{K}\cdots \sum_{i_{K(n_r-n_s)}=1}^{K}  \frac{m_1!\cdots m_K!}{\lambda_1^{m_1}\cdots \lambda_K^{m_K} } \biggr) \\& \cdot \left (  \frac{\gamma}{4P} \right ) ^{-\sum_{i=1}^{K}m_i+1} \left (  \frac{\gamma}{4P} \right ) ^{-K(n_s-1)}.
\label{PEPM18}
\end{split}
\end{equation}
Finally, substituting $\sum_{i=1}^{K}m_i=K(n_r-n_s)$ in (\ref{PEPM18}), we get
\begin{multline}
P(\mathbf {\tilde{X}}_i \rightarrow \mathbf {\tilde{X}}_j)  \leq \frac{ \rho \cdot n_r!}{(n_r-n_s)!(n_s-1)! (K!)^{n_r-n_s}}  \\ \cdot \frac{1}{\left(\prod_{i=1}^{K}\lambda_i \right)^{n_s}} \cdot\biggl(  \sum_{i_1=1}^{K}\cdots \sum_{i_{K(n_r-n_s)}=1}^{K}  \frac{m_1!\cdots m_K!}{\lambda_1^{m_1}\cdots \lambda_K^{m_K} } \biggr) \\ \cdot \left (  \frac{\gamma}{4P} \right ) ^{-Kn_r}. 
\label{PEPM19}
\end{multline}
Note that the inequality (\ref{PEPM19}) implies that a diversity order of $n_rK$ ($=n_rn_tP$) is achieved in a full rank multi-antenna OTFS system when $n_s$ antennas are selected at the receiver. We can now specialize the above diversity result for the considered multi-antenna OTFS systems which are full rank as follows.  
\begin{itemize}
\item SIMO-OTFS systems without phase rotation for $P=1$ and with phase rotation for $P>1$ are full rank. Therefore, in these cases, full spatial and DD diversity of $n_rP$ is achieved when $n_s$ receive antennas are selected. 
\item STC-OTFS systems with Alamouti code without phase rotation for $P=1$ and with phase rotation for $P>1$ are also full rank. Therefore, in these cases, full spatial and DD diversity of $2n_rP$ is achieved when $n_s$ received antennas are selected.
\end{itemize}
The above diversity results have been summarized in Table \ref{sumres}.

\subsection{Rank deficient multi-antenna OTFS systems with RAS}
\label{sec3b}
Consider the case of rank deficient multi-antenna OTFS systems with receive antenna selection. Let $\mathbf{\tilde{X}}_i$ and $\mathbf{\tilde{X}}_j$ be two distinct symbol matrices. Let $r < K$ be the minimum rank of $(\mathbf{\tilde{X}}_i-\mathbf{\tilde{X}}_j)$. For rank deficient case, the diversity analysis follows from (\ref{PEPM6})-(\ref{PEPM16}), except now $\lambda_1,\cdots,\lambda_r > 0$, $\lambda_{r+1}=,\cdots,=\lambda_{K}=0$.  Therefore, in the high SNR regime, the average PEP between $\mathbf{\tilde{X}}_i$ and $\mathbf{\tilde{X}}_j$, assuming 
$\mathbf{\tilde{X}}_i$ to be the transmitted symbol matrix, is given by
\begin{multline}
P(\mathbf {\tilde{X}}_i \rightarrow \mathbf {\tilde{X}}_j)  \leq   \frac{ \rho \cdot n_r!}{(n_r-n_s)!(n_s-1)! (K!)^{n_r-n_s}}   \\ \cdot \frac{1}{\left(\prod_{i=1}^{r}\lambda_i \right)^{n_s}} 
\cdot \biggl[  \sum_{i_1=1}^{K}\cdots \sum_{i_{K(n_r-n_s)}=1}^{K}  \\ \frac{m_1!\cdots m_{K}!}{\lambda_1^{m_1}\cdots \lambda_r^{m_r} }  \left (  \frac{\gamma}{4P} \right ) ^{-{\sum_{i=1}^{r}m_i}} \biggr]  \cdot \left(\frac{\gamma}{4P} \right)^{-rn_s}.
\label{PEPM21}
\end{multline}
Since $\sum_{i=1}^{K}m_i=K(n_r-n_s)$, it follows that $0 \leq \sum_{i=1}^{r}m_i \leq K(n_r-n_s)$. It is observed that the term in the square brackets is function of $\frac{\gamma}{4P}$ and there exist terms $i_1 \cdots i_{K(n_r-n_s)}$ such that $\sum_{i=1}^{r}m_i=0$. Regrouping the terms in (\ref{PEPM21}), we can write
\begin{equation} 
\begin{split}
P(\mathbf {\tilde{X}}_i \rightarrow \mathbf {\tilde{X}}_j)  &\leq \frac{\rho \cdot n_r!}{(n_r-n_s)!(n_s-1)! (K!)^{n_r-n_s}}    \frac{1}{\left(\prod_{i=1}^{r}\lambda_i \right)^{n_s}}  \\&   \left ( \sum_{j=0}^{K(n_r-n_s)}\psi_j \left(\frac{\gamma}{4P}\right)^{-j} \right )  \cdot \left ( \frac{\gamma}{4P} \right )^{-rn_s},
\label{PEPM22}
\end{split}
\end{equation}
where $j=\sum_{i=1}^{r}m_i$ and $\psi_j$ is the sum of the terms multiplying $\left ( \frac{\gamma}{4P} \right )^{-\sum_{i=}^{r}m_i}$ with the same exponents. For sufficiently high SNRs, the term $\left ( \frac{\gamma}{4P} \right )^{-j}$ vanishes for $\sum_{i=1}^{r}m_i >0$. Thus, we have
\begin{equation} 
\begin{split}
P(\mathbf {\tilde{X}}_i \rightarrow \mathbf {\tilde{X}}_j)  &\leq \frac{\rho \cdot n_r!}{(n_r-n_s)!(n_s-1)! (K!)^{n_r-n_s}} \\ & \cdot \frac{1}{\left(\prod_{i=1}^{r}\lambda_i \right)^{n_s}}   \psi_0  \cdot \left ( \frac{\gamma}{4P} \right )^{-rn_s}.
\label{PEPM23}
\end{split}
\end{equation}
The above expression shows that a diversity order of $n_sr$ is achieved for a rank deficient multi-antenna OTFS system when $n_s$ antennas are selected at the receiver. We specialize the above diversity result for the considered multi-antenna OTFS systems which are rank deficient as follows.
\begin{itemize}
\item The minimum rank of $(\mathbf{\tilde{X}}_i-\mathbf{\tilde{X}}_j)$ is $1$ for SIMO-OTFS ($P>1$) and MIMO-OTFS ($P\geq 1$) systems without phase rotation. Therefore, these systems achieve a diversity of $n_s$ when $n_s$ antennas are selected at the receiver. 
\item For MIMO-OTFS $(P\geq 1)$ systems with phase rotation, the minimum rank of $(\mathbf{\tilde{X}}_i-\mathbf{\tilde{X}}_j)$ is $P$, which is rank deficient. Therefore, these systems achieve a diversity of $n_sP$ when $n_s$ receive antennas are selected. 
\item For STC-OTFS ($P>1$) systems with Alamouti code without phase rotation, the minimum rank of $(\mathbf{\tilde{X}}_i-\mathbf{\tilde{X}}_j)$ is $2$, which is rank deficient. Therefore, these systems achieve a diversity of $2n_s$ when $n_s$ receive antennas are selected.
\end{itemize}
The above diversity results have been summarized in Table \ref{sumres}.

\begin{table}
\centering
\begin{tabular}{|l|c|c|c|c|} 
\hline
\multirow{2}{*}{OTFS system} &  \multirow{2}{*}{\begin{tabular}[c]{@{}l@{}}\# ant. \\ selected \end{tabular}} & \# DD & \multicolumn{2}{c|}{Diversity order} \\ \cline{4-5}
& & paths &  without PR & with PR  \\ 
\hline 
SIMO-OTFS,    & $n_s\geq 1$ & $P=1$ & $n_r$ & $n_r$ \\ \cline{3-5}
$n_r\geq 1$                  & & $P > 1$ & $n_s$ & $n_rP$ \\ \hline
MIMO-OTFS,  & $n_s\geq n_t$  & $P \geq 1$  & $n_s$  & $n_sP$   \\ 
$n_r\geq n_t$                & & & & \\ \hline
STC-OTFS (Alamouti) & $n_s \geq 1$ & $P=1$ & $2n_r$ & $2n_r$ \\ \cline{3-5}
$n_t=2$, $n_r\geq1$    & & $P>1$ & $2n_s$ & $2n_rP$ \\ \hline
\end{tabular}
\caption{Summary of diversity order results for multi-antenna OTFS systems with RAS.}
\vspace{-5mm}
\label{sumres}
\end{table}

\section{Simulation results}
\label{sec5}
In this section, we present simulation results on the bit error performance that validate the analytical diversity results derived in the previous section. We evaluate the bit error rate (BER) of the considered multi-antenna OTFS systems without and with phase rotation for $P=1,2,4$ and $n_s \geq 1$. The simulation parameters used are listed in Table \ref{SimPar}.

{\em SIMO-OTFS (without phase rotation) for $P=1$:} 
Figure \ref{Sel_BER1} shows the simulated BER performance of SIMO-OTFS without phase rotation for $P=1$, $M=N=2$, $n_s=1$, $n_r=1,2,3,4$, BPSK, and ML detection. A carrier frequency of 4 GHz, subcarrier spacing of 3.75 kHz, and a maximum speed of 506.2 km/h are considered. The considered carrier frequency and maximum speed correspond to a maximum Doppler of 1.875 kHz. The DD channel model is as per (\ref{sparsechannel}) and the DD profiles for different values of $P$ are presented in Table \ref{SimPar}. The considered system is full rank and the analytically predicted diversity order is $n_r$ (refer Table \ref{sumres} and Sec. \ref{sec3a}). The BER plots in Fig. \ref{Sel_BER1} show that the system indeed achieves first, second, third, and fourth order diversity slopes for $n_r=1,2,3,$ and 4, respectively, corroborating the analytically predicted diversity orders.

\begin{table}[t]
\begin{center}
\begin{tabular}{|l|l|} \hline
\textbf{Parameter} & \textbf{Value} \\ \hline
Carrier frequency, $f_c$ &  \\ 
(GHz) & 4 \\ \hline
Subcarrier spacing, $\Delta f$ &\\ (kHz) &  3.75 \\ \hline
DD profile for $P=1$ & \\ ($\tau_i$ (sec), $\nu_i$ (Hz))&   $(\frac{1}{M\Delta f}$, $\frac{1}{NT})$\\ \hline
DD profile for $P=2$ & \\ $\&$ $M=2,4$, $N=2$ & $(0, 0)$, $(\frac{1}{M\Delta f}$, $\frac{1}{NT})$\\ \hline
DD profile for $P=2$ & \\ $\&$ $M=4$, $N=4$ & $(\frac{1}{M\Delta f}$, $\frac{1}{NT})$, $(\frac{2}{M\Delta f}$, $\frac{2}{NT})$ \\ \hline
DD profile for $P=4$ & \\  $\&$ $M=2$, $N=2$ & $(0, 0)$, $(0, \frac{1}{NT})$, $(\frac{1}{M\Delta f}, 0)$, $(\frac{1}{M\Delta f}$, $\frac{1}{NT})$\\ \hline
Maximum speed (km/h) & 506.2 \\ \hline
Modulation scheme & BPSK, 16-QAM \\ \hline
\end{tabular}
\caption{Simulation parameters.}
\vspace{-6mm}
\label{SimPar}
\end{center}
\end{table} 

\begin{figure}[t]
\includegraphics[width=9.0cm, height=6.0cm]{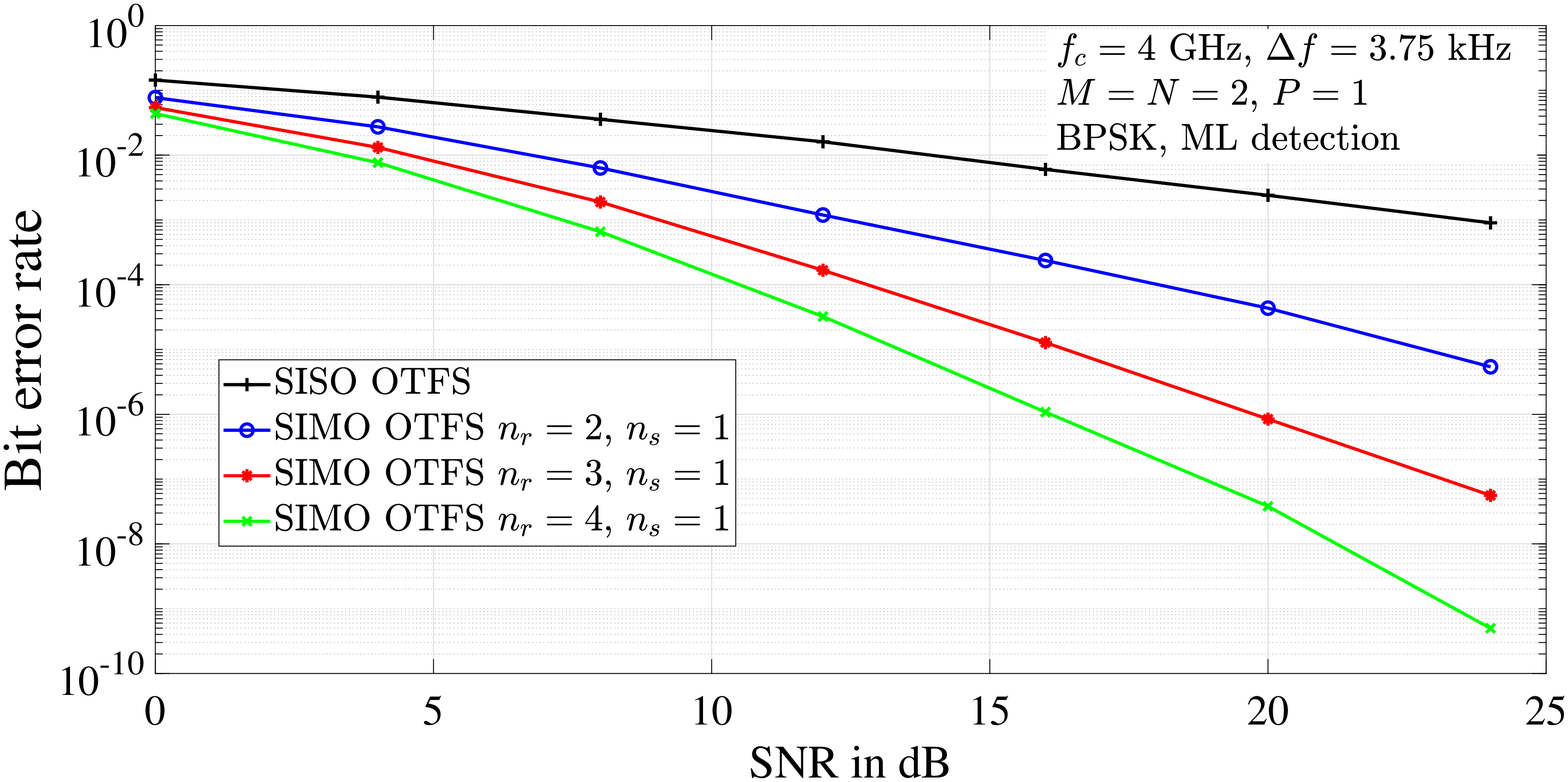}
\vspace{-2mm}
\caption{BER performance of SIMO-OTFS without phase rotation for $P=1$, $M=N=2$, $n_s=1$, and $n_r=1,2,3,4$.}
\vspace{-2mm}
\label{Sel_BER1}
\end{figure}

{\em SIMO-OTFS (without phase rotation) for $P>1$:} 
Figure \ref{Sel_BER2} shows the simulated BER performance of SIMO-OTFS without phase rotation for $P=4$, $M=N=2$, $n_s=1$, $n_r=1,4$, BPSK, and ML detection. Other simulation parameters are as given in Table \ref{SimPar}. In addition to the simulated BER plot, upper bound and lower bounds on the bit error performance are also plotted. The upper bound on the bit error probability is obtained from PEP using union bound, as 
\begin{equation}
P_b \leq \frac{1}{L n_tMN\log_2|\mathbb{A}|}\sum_{i=1}^{L}\sum_{j=1,j\neq i}^{L}P(\tilde{\mathbf{X}}_i \rightarrow \tilde{\mathbf{X}}_j),
\label{BER}
\end{equation}
where $L=|\mathbb{A}^{n_tMN}|$. The lower bound is obtained based on summing the PEPs corresponding to all the pairs $\mathbf{X}_i$ and $\mathbf{X}_j$ such that the difference matrix $(\mathbf{X}_i-\mathbf{X}_j)$ has rank one \cite{otfs_div}. The considered system is rank deficient and the analytically predicted diversity order is $n_s$ (refer Sec. \ref{sec3b} and Table \ref{sumres}). Since the number of antennas selected is $n_s=1$, the predicted diversity order is 1. We can make two key observations from Fig. \ref{Sel_BER2}. First, the diversity slope is one for both $n_r=1$ and $n_r=4$. Second, The upper bound, lower bound, and simulated BER almost merge at high SNRs. These observations validate the simulation results as well the analytically predicted diversity order.

{\em SIMO-OTFS (without and with phase rotation) for $P>1$:}
Figure \ref{sel_BER3} shows the BER performance of SIMO-OTFS without and with phase rotation for $P=2$, $M=N=4$, $n_s=1$, $n_r=1,2$, BPSK, ML detection, and other parameters as in Table \ref{SimPar}. For $P>1$, SIMO-OTFS without phase rotation is rank deficient and the analytical diversity order is $n_s$. With phase rotation, the system is full-ranked and it has a diversity order of $n_rP$ (refer Sec. \ref{sec3b}, Sec. \ref{sec3a}, and Table \ref{sumres}). For the considered system, the predicted diversity orders are 1 and 4 for without and with phase rotation, respectively. The slopes in the BER plots in Fig. \ref{sel_BER3} are observed to be in line with the predicted diversity orders.

{\em SIMO-OTFS (without and with phase rotation) for 16-QAM:}
Figure \ref{sel_BER31} shows the BER performance of SIMO-OTFS without and with phase rotation for 16 QAM, $P=2$, $M=N=2$, $n_s=1$, $n_r=1,2$, ML detection, and other parameters as in Table \ref{SimPar}. For $P>1$, the analytically predicted diversity orders for the considered SIMO-OTFS system without and with phase rotation are 1 ($n_s$) and 4 ($n_rP$), respectively. In Fig. \ref{sel_BER31}, the diversity slopes are found to follow these diversity orders.

\begin{figure}
\includegraphics[width=9cm, height=6cm]{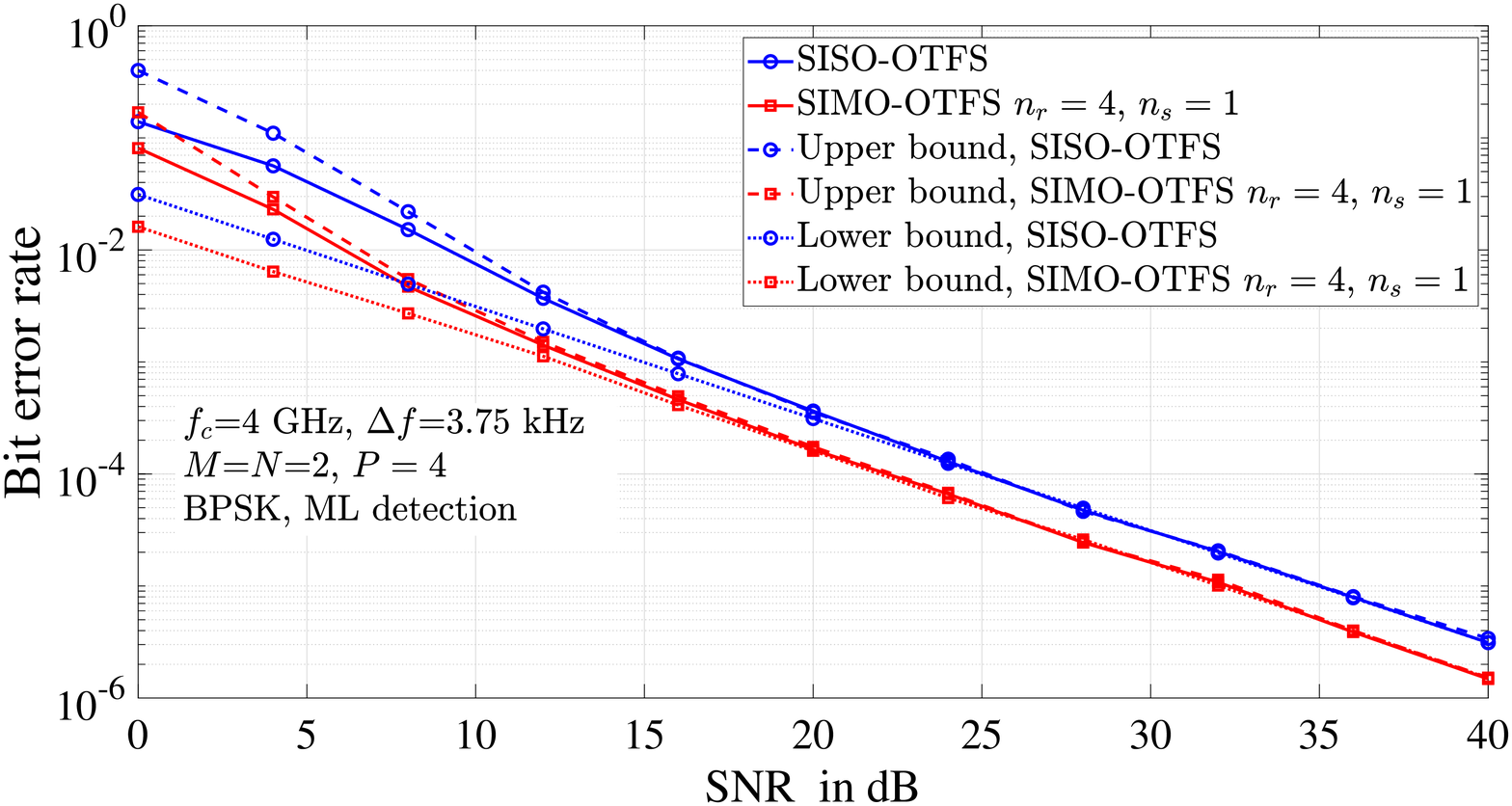}
\vspace{-2mm}
\caption{BER performance of SIMO-OTFS without phase rotation for $P=4$, $M=N=2$, $n_s=1$, and $n_r=1,4$.}
\vspace{-4mm}
\label{Sel_BER2}
\end{figure}

\begin{figure}
\includegraphics[width=9cm, height=6cm]{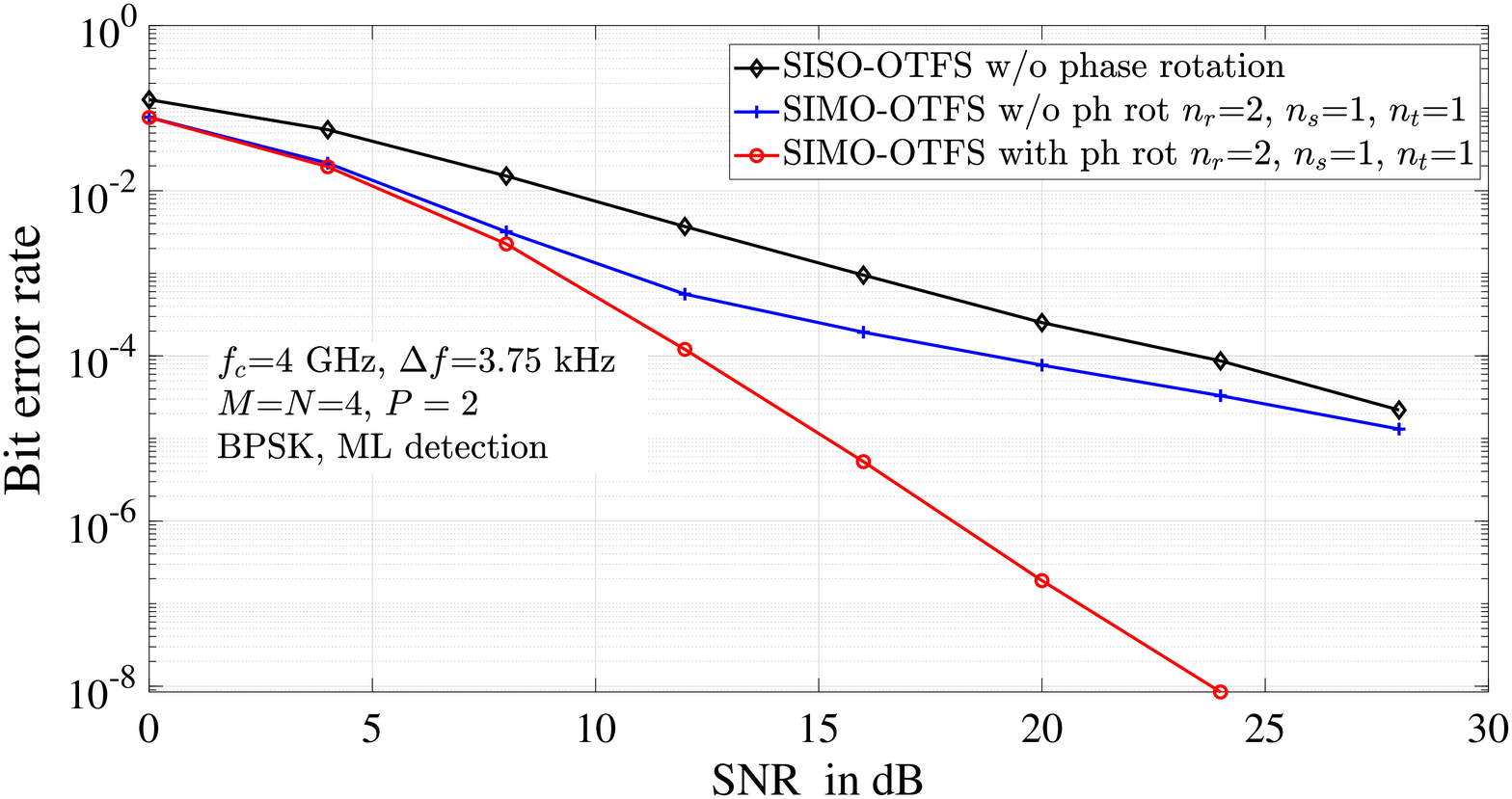}
\vspace{-2mm}
\caption{BER performance of SIMO-OTFS without and with phase rotation for $P=2$, $M=N=4$, $n_s=1$, and $n_r=1,2$.}
\vspace{-4mm}
\label{sel_BER3}
\end{figure}

\begin{figure}
\includegraphics[width=9cm, height=6cm]{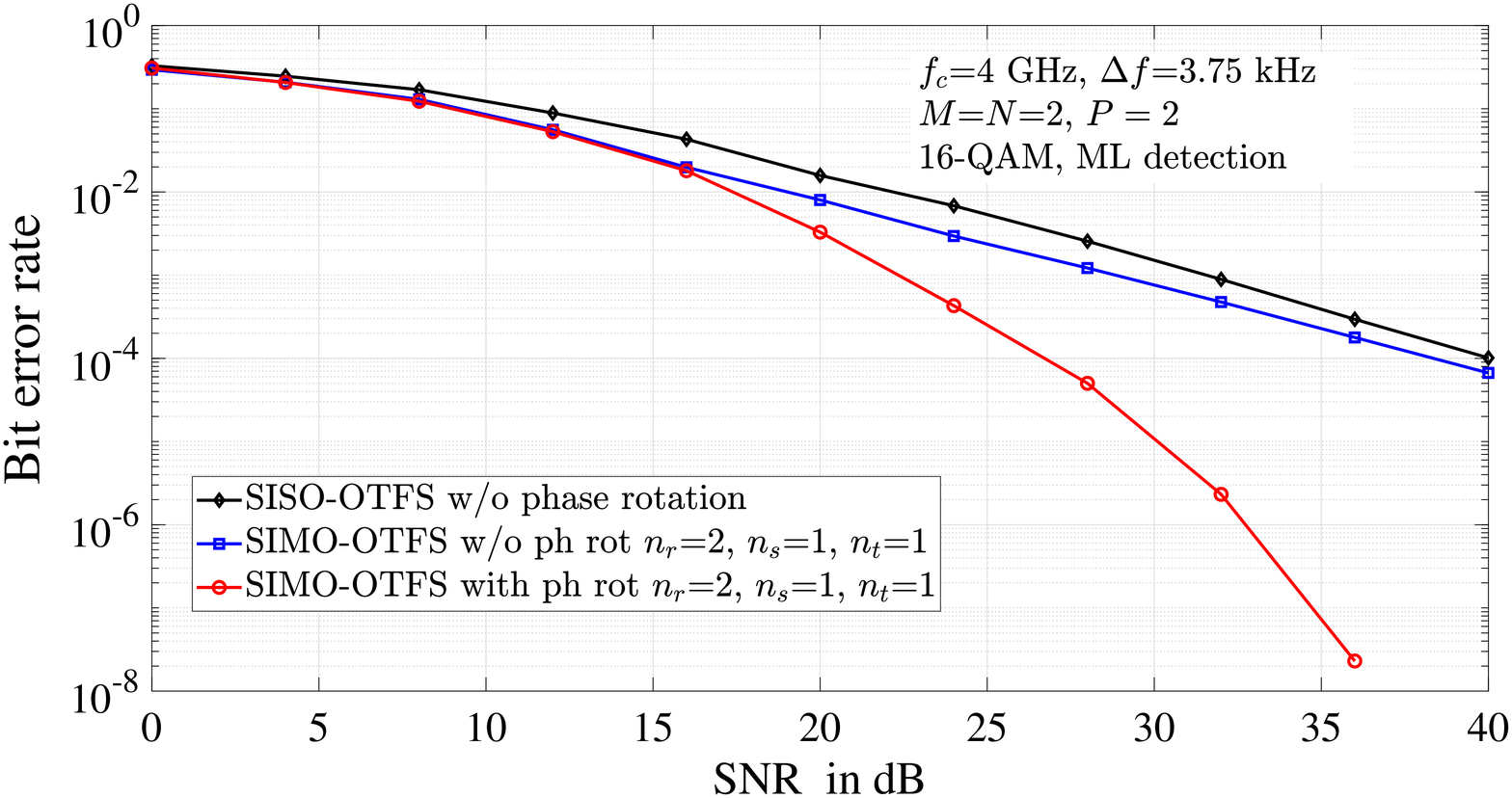}
\vspace{-2mm}
\caption{BER performance of SIMO-OTFS without and with phase rotation for $P=2$, $M=N=2$, $n_s=1$, $n_r=1,2$, and 16-QAM.}
\vspace{-4mm}
\label{sel_BER31}
\end{figure}

{\em Alamouti STC-OTFS (without and with phase rotation) for $P>1$:} 
Figure \ref{Sel_BER5} shows the BER performance of Alamouti STC-OTFS without phase rotation for $P=2$, $M=N=2$, $n_t=2$, $n_s=1,2$, $n_r=1,2,3$, BPSK, ML detection, and other parameters as in Table \ref{SimPar}. From Fig. \ref{Sel_BER5}, it is observed that the achieved diversity order is 2 for $n_s=1$ and 4 for $n_s=2$. This corroborates with the predicted diversity order of $2n_s$, the system being rank deficient. For the above Alamouti STC-OTFS system, Fig. \ref{Sel_BER6} shows the performance with phase rotation. This system with phase rotation is full-ranked with a predicted diversity order of $2n_rP$. The diversity slopes observed in Fig. \ref{Sel_BER6} are in accordance with this analytical prediction.

\begin{figure}
\includegraphics[width=9cm, height=6cm]{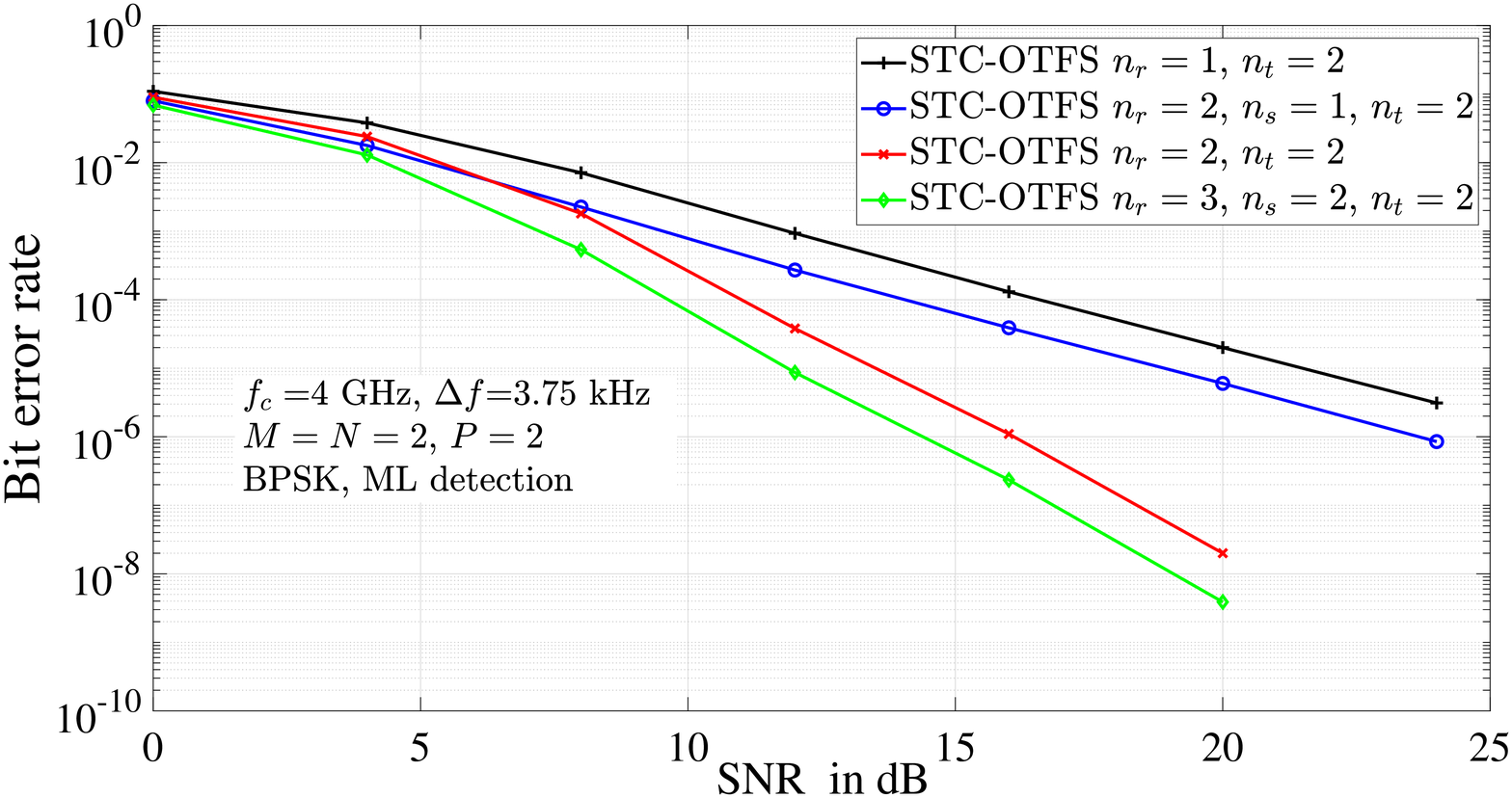}
\vspace{-2mm}
\caption{BER performance Alamouti STC-OTFS without phase rotation for $P=2$, $M=N=2$, $n_t=2$, $n_s=1,2$, and $n_r=1,2,3$.}
\vspace{-4mm}
\label{Sel_BER5}
\end{figure}

\begin{figure}[t]
\includegraphics[width=9cm, height=6cm]{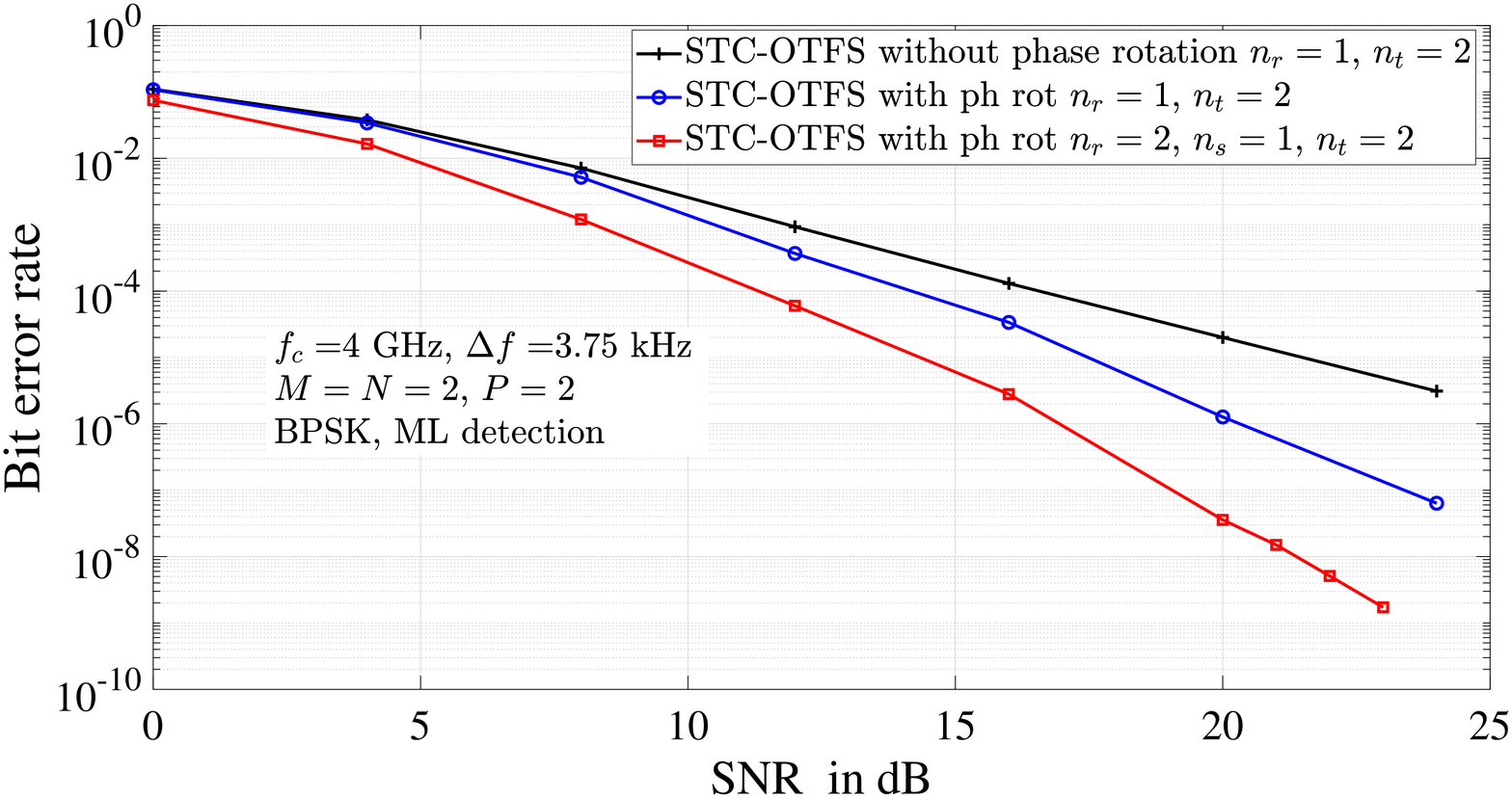}
\vspace{-2mm}
\caption{BER performance of Alamouti STC-OTFS with phase rotation for $P=2$, $M=N=2$, $n_t=2$, $n_s=1$, and $n_r=1,2$.}
\vspace{-4mm}
\label{Sel_BER6}
\end{figure}

\begin{figure}[t]
\includegraphics[width=9cm,height=6cm]{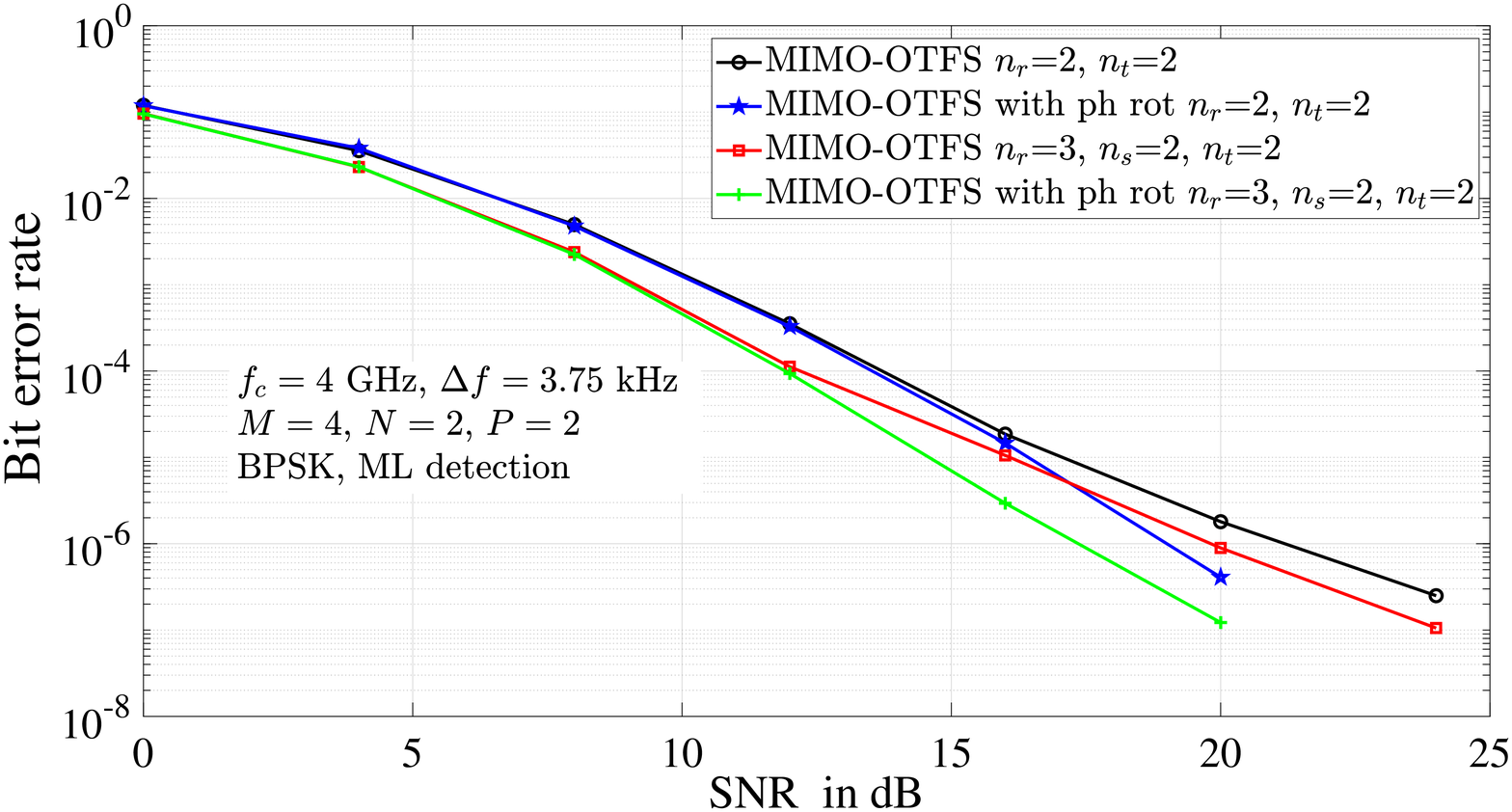}
\vspace{-3mm}
\caption{BER performance of MIMO-OTFS without and with phase rotation for $P=2$, $M=4$, $N=2$, $n_t=2$, $n_s=2$, and $n_r=2,3$.}
\vspace{-4mm}
\label{Sel_BER8}
\end{figure}

{\em MIMO-OTFS (without and with phase rotation) for $P>1$:}
Figure \ref{Sel_BER8} shows the BER performance of MIMO-OTFS without and with phase rotation for $P=2$, $M=4$, $N=2$, $n_t=2$, $n_s =2$, $n_r=2,3$, BPSK, and other parameters as in Table \ref{SimPar}. The considered systems are rank deficient, and the predicted diversity orders are $n_s$ and $n_sP$ for without and with phase rotation, respectively. It can be seen in Fig. \ref{Sel_BER8} that, as predicted, MIMO-OTFS without phase rotation achieves 2nd order diversity slope and with phase rotation achieves 4th order diversity slope.

\section{Conclusions}
\label{sec6}
We analyzed the diversity performance of receive antenna selection in multi-antenna OTFS systems. Antennas were selected based on the maximum channel Frobenius norms in the DD domain. Our diversity analysis results showed that, with no phase rotation, SIMO-OTFS and MIMO-OTFS systems with RAS are rank deficient, and therefore they do not extract the full receive diversity as well as the diversity present in the DD domain. Also, Alamouti coded STC-OTFS system with RAS and no phase rotation was shown to extract the full transmit diversity, but it failed to extract the DD diversity. On the other hand, SIMO-OTFS and STC-OTFS systems with RAS become full-ranked when phase rotation is used, because of which they extracted the full spatial as well as the DD diversity present in the system. When phase rotation is used, MIMO-OTFS systems with RAS was shown to extract the full DD diversity, but they did not extract the full receive diversity because of rank deficiency. Detailed simulation results validated the analytically predicted diversity performance.

\appendix[Analysis for fractional Delays and Dopplers ]
\label{sec7}
\subsection{Input-Output relation with fractional delays and Dopplers}
\label{secap1}
Considering the channel representation in DD defined in (\ref{sparsechannel}) with non-zero fractional delays and Dopplers, we have
\begin{equation} 
\tau_i=\frac{\alpha_i+a_i }{M\Delta f}, \ \ \nu_i=\frac{\beta_i+b_i }{NT},
\label{ap1}
\end{equation} 
where $\alpha_i=[\tau_iM\Delta f]^\odot$, $\beta_i=[\nu_iNT]^\odot$, $[.]^\odot$ denotes the rounding operator (nearest integer), $\alpha_i$, $\beta_i$ are assumed to be integers corresponding to the indices of the delay tap and Doppler frequency associated with $\tau_i$ and $\nu_i$, respectively, and $a_i$, $b_i$ are the fractional delay and Doppler satisfying $-\frac{1}{2}<a_i,b_i \leq \frac{1}{2}$. The DD channel with fractional delays and Dopplers, assuming rectangular window functions, can be written as
\begin{equation} 
h(\tau,\nu)=\sum_{i=1}^{P}h_i e^{-j2\pi\tau_i\nu_i}\mathcal{G}(\nu,\nu_{i})\mathcal{F}(\tau,\tau_i),
\label{ap2}
\end{equation} 
where
\begin{equation}
\begin{split} 
\mathcal {G}(\nu,\nu _{i})\triangleq&\sum _{n'=0}^{N-1} e^{-j2\pi (\nu -\nu _{i})n'T}, \\ \mathcal {F}(\tau,\tau _{i})\triangleq&\sum _{m'=0}^{M-1} e^{j2\pi (\tau -\tau _{i})m' \Delta f}. 
\label{ap3}
\end{split}
\end{equation}
The input-output relation with fractional delay-Doppler can be written as \cite{otfs_div}
\begin{equation} 
\begin{split} 
y[k,l]=&\sum _{i=1}^{P}\sum _{q=0}^{M-1}\sum _{q'=0}^{N-1} \left ({\frac {e^{j2\pi (-q-a_{i})}-1}{Me^{j\frac {2\pi }{M}(-q-a_{i})}-M}}\right) \\&\cdot \left ({\frac {e^{-j2\pi (-q'-b_{i})}-1} {Ne^{-j\frac {2\pi }{N}(-q'-b_{i})}-N}}\right) h_{i}e^{-j2\pi \tau _{i} \nu _{i}} \\&\cdot \,x[(k-\beta _{i}+q')_{N},(l-\alpha _{i}+q)_{M}]. 
\label{ap4}
\end{split} 
\end{equation}
Vectorizing the input-output relation in (\ref{ap4}), we can write 
\begin{equation} 
\mathbf {y}=\mathbf {Hx}+\mathbf {v},
\label{ap5}
\end{equation}
where $\mathbf{y}$ $\in \mathbb{C}^{MN \times 1}$ is the received signal vector, $\mathbf{x}$ $\in \mathbb{C}^{MN \times 1}$ transmit signal vector, $\mathbf{H}$ $\in \mathbb{C}^{MN \times MN}$ is the equivalent channel matrix, and $\mathbf{v}$ $\in \mathbb{C}^{MN \times 1}$ is the noise vector.

Based on (\ref{ap4}), the input-output relation with receive antennas selection in (\ref{mimosel2}) can be extended to fractional delays and Dopplers, as
\begin{equation}
\mathbf{\bar{y}}'=\mathbf{\bar{H}}'\mathbf{\bar{x}}+\mathbf{\bar{v}}',
\label{ap8}
\end{equation}
where $\mathbf{\bar{y}'} \in \mathbb{C}^{{n_s}MN \times 1}$ is the received signal vector, $\mathbf{\bar{H}'} \in \mathbb{C}^{n_{s}MN\times n_tMN}$ is the channel matrix with antenna selection, $\mathbf{\bar{x}} \in \mathbb{C}^{n_tMN \times 1}$ is the OTFS transmit vector, and  $\mathbf{\bar{v}'} \in \mathbb{C}^{{n_s}MN \times 1}$ is the noise vector.

\subsection{Diversity analysis for $P=1$}
\label{secap3}
The selection rule in (\ref{selr1}) and (\ref{selr2}) are equivalent for $P=1$. Therefore, for diversity analysis for $P=1$, the input-output relation in (\ref{ap8}) can be written in an alternate form as
\begin{equation}
\mathbf{\tilde{Y}}=\mathbf{\tilde{H}}\mathbf{\tilde{X}}+\mathbf{\tilde{V}},
\label{ap9}
\end{equation}
where $\mathbf{\tilde{Y}}\in \mathbb{C}^{n_s \times MN}$ with its $i$th row corresponding to the received signal in the $i$th selected receive antenna, $\mathbf{\tilde{H}} \in \mathbb{C}^{n_s \times n_t}$ is the channel matrix whose $(i,j)$th element is $h_{ij}e^{-j2\pi \tau \nu}$, $\mathbf{\tilde{X}}$ is $n_t \times MN$ symbol matrix whose $i$th column $\mathbf{\tilde{X}}[i]$ is given by (\ref{ap7}) shown at the top of next page, and $\mathbf{\tilde{V}} \in \mathbb{C}^{n_s \times MN}$ is the noise matrix.
\begin{table*}
\normalsize
\centering
\begin{align}
\mathbf{X}[i]=\begin{bmatrix}
\sum _{q=0}^{M-1}\sum _{q'=0}^{N-1} \left ({\frac {e^{j2\pi (-q-a)}-1}{Me^{j\frac {2\pi }{M}(-q-a)}-M}}\right) \ \left ({\frac {e^{-j2\pi (-q'-b)}-1} {Ne^{-j\frac {2\pi }{N}(-q'-b)}-N}}\right)x_{1}[(k-\beta+q')_N,(l-\alpha+q)_M]\\ 
\sum _{q=0}^{M-1}\sum _{q'=0}^{N-1} \left ({\frac {e^{j2\pi (-q-a)}-1}{Me^{j\frac {2\pi }{M}(-q-a)}-M}}\right) \ \left ({\frac {e^{-j2\pi (-q'-b)}-1} {Ne^{-j\frac {2\pi }{N}(-q'-b)}-N}}\right)x_{2}[(k-\beta+q')_N,(l-\alpha+q)_M]\\ 
\vdots \\ 
\sum _{q=0}^{M-1}\sum _{q'=0}^{N-1} \left ({\frac {e^{j2\pi (-q-a)}-1}{Me^{j\frac {2\pi }{M}(-q-a)}-M}}\right) \ \left ({\frac {e^{-j2\pi (-q'-b)}-1} {Ne^{-j\frac {2\pi }{N}(-q'-b)}-N}}\right)x_{n_t}[(k-\beta+q')_N,(l-\alpha+q)_M]\\ 
\end{bmatrix}.
\label{ap7}
\end{align}
\medskip
\hrule
\end{table*}

\subsubsection{Full rank case}
\label{secap3a}
Let $\mathbf{\tilde{X}}_i$ and $\mathbf{\tilde{X}}_j$ be two distinct symbol matrices. The conditional PEP between $\mathbf{\tilde{X}}_i$ and $\mathbf{\tilde{X}}_j$, assuming perfect DD channel knowledge and ML detection, is given by
\begin{equation} 
P(\mathbf {\tilde{X}}_i \rightarrow \mathbf {\tilde{X}}_j|\mathbf {\tilde{H}},\mathbf {\tilde{X}}_i)=Q \left ({\sqrt {\frac {\|\mathbf {\tilde{H}}(\mathbf {\tilde{X}}_i-\mathbf {\tilde{X}}_j)\|^{2}}{2N_0}} }\right)\!.
\label{ap10}
\end{equation}
Upper bounding (\ref{ap10}) using Chernoff bound and averaging over the distribution of $\mathbf{\tilde{H}}$, the unconditional PEP can be written as
\begin{equation} 
P(\mathbf {\tilde{X}}_i\rightarrow \mathbf {\tilde{X}}_j) \leq \mathbb {E}_{\mathbf{\tilde{H}}} \left [{ {\exp} \left (-{{\frac {\gamma ~\|\mathbf {\tilde{H}} (\mathbf {\tilde{X}}_i-\mathbf {\tilde{X}}_j)\|^{2}}{4}}\ }\right) }\right]\!.
\label{ap11}
\end{equation} 
The distribution of $\mathbf{\tilde{H}}$ is given in (\ref{distm}). Therefore, the PEP can be written as
\begin{equation} 
\begin{split}
P(\mathbf {\tilde{X}}_i \rightarrow \mathbf {\tilde{X}}_j) &\leq \sum_{l=1}^{n_s}\int_{\mathcal{\tilde{H}}_l}^{}e^{\frac{-\gamma}{4}\|\mathbf{\tilde{H}}(\mathbf{\tilde{X}}_i-\mathbf{\tilde{X}}_j)\|^2}  \frac{n_r!}{(n_r-n_s)!n_s!}\\ & \hspace{-10mm} \cdot \left ( 1-e^{-\|\mathbf{h}'_l\|^2} \sum_{k=0}^{n_t-1} \frac{\|\mathbf{h}'_l\|^{2k}}{k!}\right )^{n_r-n_s} \\ & \hspace{-10mm} \cdot \frac{1}{\pi ^{n_sn_t}}e^{-(\|\mathbf{h}'_1\|^2+\cdots+\|\mathbf{h}'_{n_s}\|^2)}d\mathbf{h}'_1 \cdots d\mathbf{h}'_{n_s}.
\label{ap12}
\end{split}
\end{equation} 
Following the steps from (\ref{PEPM4})-(\ref{PEPM19}) in Sec. \ref{sec3a}, we can write PEP as
\begin{multline}
P(\mathbf {\tilde{X}}_i \rightarrow \mathbf {\tilde{X}}_j)  \leq \frac{   n_r!}{(n_r-n_s)!(n_s-1)! (n_t!)^{n_r-n_s}}   \\ \cdot \frac{1}{\left(\prod_{i=1}^{n_t}\lambda_i \right)^{n_s}}  \cdot\biggl(  \sum_{i_1=1}^{n_t}\cdots \sum_{i_{n_t(n_r-n_s)}=1}^{n_t}  \frac{m_1!\cdots m_{n_t}!}{\lambda_1^{m_1}\cdots \lambda_{n_t}^{m_{n_t}} } \biggr)  \\ \cdot \left (  \frac{\gamma}{4} \right ) ^{-n_tn_r}. 
\label{ap13}
\end{multline}
The above equation shows that, for fractional delay-Doppler also, diversity of $n_rn_t$ is achieved when $n_s$ antennas are selected at the receiver. Therefore, full spatial diversity is achieved for a full rank multi-antenna OTFS system. We can specialize the above generalized result for multi-antenna OTFS systems which are full rank for $P=1$, as follows.
\begin{itemize}
\item SIMO-OTFS system for $P=1$ is full rank. Therefore, for this system, full spatial diversity of $n_r$ is achieved when $n_s$ antennas are selected at the receiver. 
\item STC-OTFS system with Alamouti code for $P=1$ is also full rank. Therefore, this system also achieves full spatial diversity of $2n_r$ when $n_s$ receive antennas are selected.
\end{itemize}

\subsubsection{Rank deficient case}
\label{secap3b}
Let $\mathbf{\tilde{X}}_i$ and $\mathbf{\tilde{X}}_j$ be two distinct symbol matrices. Let $r$ be the minimum rank of $(\mathbf{\tilde{X}}_i$-$\mathbf{\tilde{X}}_j)$  and $\lambda_1,\cdots,\lambda_r > 0$, $\lambda_{r+1}= \cdots =\lambda_{n_t}=0$ be the eigenvalues of the matrix $(\mathbf{\tilde{X}}_i-\mathbf{\tilde{X}}_j)(\mathbf{\tilde{X}}_i-\mathbf{\tilde{X}}_j)^H$. Following the diversity analysis for integer delay-Doppler in Sec. \ref{sec3b}, we can obtain the PEP expression as
\begin{eqnarray}
P(\mathbf {\tilde{X}}_i \rightarrow \mathbf {\tilde{X}}_j)  & \leq & \frac{  n_r!}{(n_r-n_s)!(n_s-1)! (n_t!)^{n_r-n_s}} \nonumber \\
& &  \frac{1}{\left(\prod_{i=1}^{r}\lambda_i \right)^{n_s}} \psi_0  \cdot \left ( \frac{\gamma}{4} \right )^{-rn_s}.
\label{ap14}
\end{eqnarray}
The above expression shows that, for the rank deficient case, diversity of $n_sr$ is achieved when $n_s$ antennas are selected at the receiver. MIMO-OTFS system with $P=1$ is rank deficient with minimum rank one. Therefore, diversity of $n_s$ is achieved when $n_s$ antennas are selected in MIMO-OTFS. 

\begin{figure}[t]
\includegraphics[width=9cm,height=6cm]{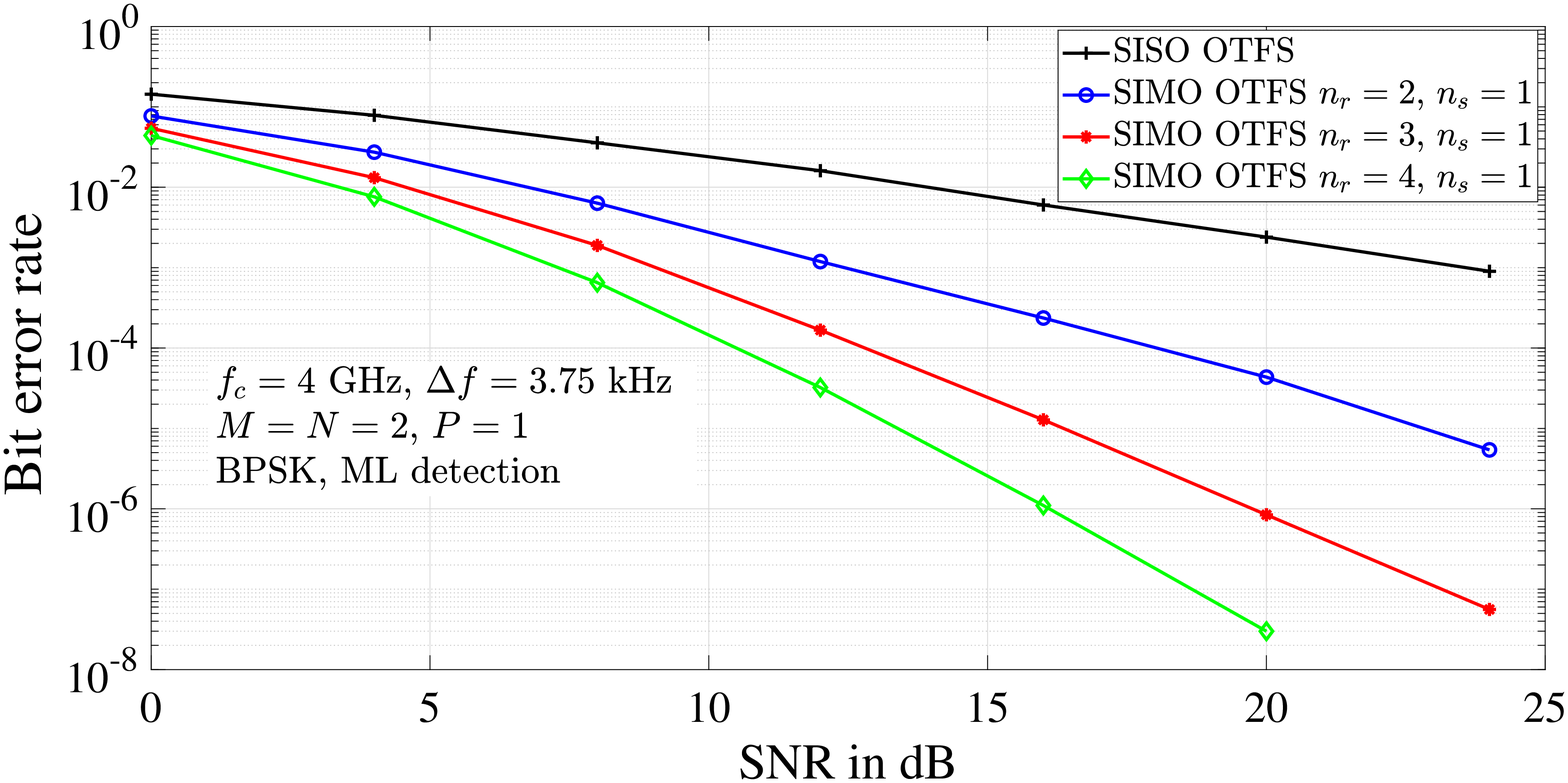}
\vspace{-2mm}
\caption{BER performance of SIMO-OTFS without phase rotation for $M=N=2$, $P=1$, $n_s=1$, and $n_r=1,2,3,4$, with fractional delay and Doppler.}
\vspace{-4mm}
\label{Sel_BER11}
\end{figure}

\subsection{Simulation results}
\label{secap4}
In this subsection, we present the simulation results for fractional delays and Dopplers. For all the simulation results presented in this subsection, the fractional delays and Dopplers are generated as follows. The Doppler shift corresponding to $i$th channel tap is generated using Jakes formula \cite{otfs4} $\nu_i=\nu_{\max}\cos(\theta_i)$, where $\nu_{\max}$ is the maximum Doppler shift  and $\theta_i$ is uniformly distributed over $[-\pi,\pi]$. The delay corresponding to $i$th channel tap is generated as  uniformly distributed over $[0,(M-1)T_s]$, where $T_s=1/(M \Delta f)$ and $\Delta f$ is the subcarrier spacing. Exponential power delay profile and Jakes Doppler spectrum are considered \cite{WC_1}.

Figure \ref{Sel_BER11} shows the simulated bit error performance of SIMO-OTFS without phase rotation for $M=2$, $N=2$, $P=1$, $n_s=1$, $n_r=1,2,3,4$, BPSK, and ML detection. The carrier frequency, subcarrier spacing, and maximum Doppler considered are 4 GHz, 3.75 kHz, and 1.875 kHz, respectively. From Fig. \ref{Sel_BER11}, it is seen that system achieves 1st, 2nd, 3rd, and 4th order diversity for $n_r=1,2,3,$ and 4, respectively, verifying analytically predicted diversity orders.

For the case of $P>1$ with fractional delays and Dopplers, the selection rule in (\ref{selr1}) and (\ref{selr2}) are not equivalent. Therefore, it is difficult to find the distribution of $\mathbf{\tilde{H}}$ because of spreading of channel coefficients in multiple DD bins, leading to intractability of analysis. Consequently, for $P>1$, we present simulation results. For this, we consider simulation parameters according to IEEE 802.11p standard for wireless access in vehicular environments (WAVE) \cite{WAVE} and long term evolution (LTE) standard \cite{lte}. Also, rectangular pulse shapes are used. Since the values of $M$ and $N$ are large, ML detection is not feasible. Therefore, we have used minimum mean square error (MMSE) detection and message passing (MP) detection \cite{otfs4}. Also, in these figures, we present a comparison between the performance of SIMO/MIMO-OTFS and SIMO/MIMO-OFDM with RAS.

\begin{figure}[t]
\includegraphics[width=9cm, height=6cm]{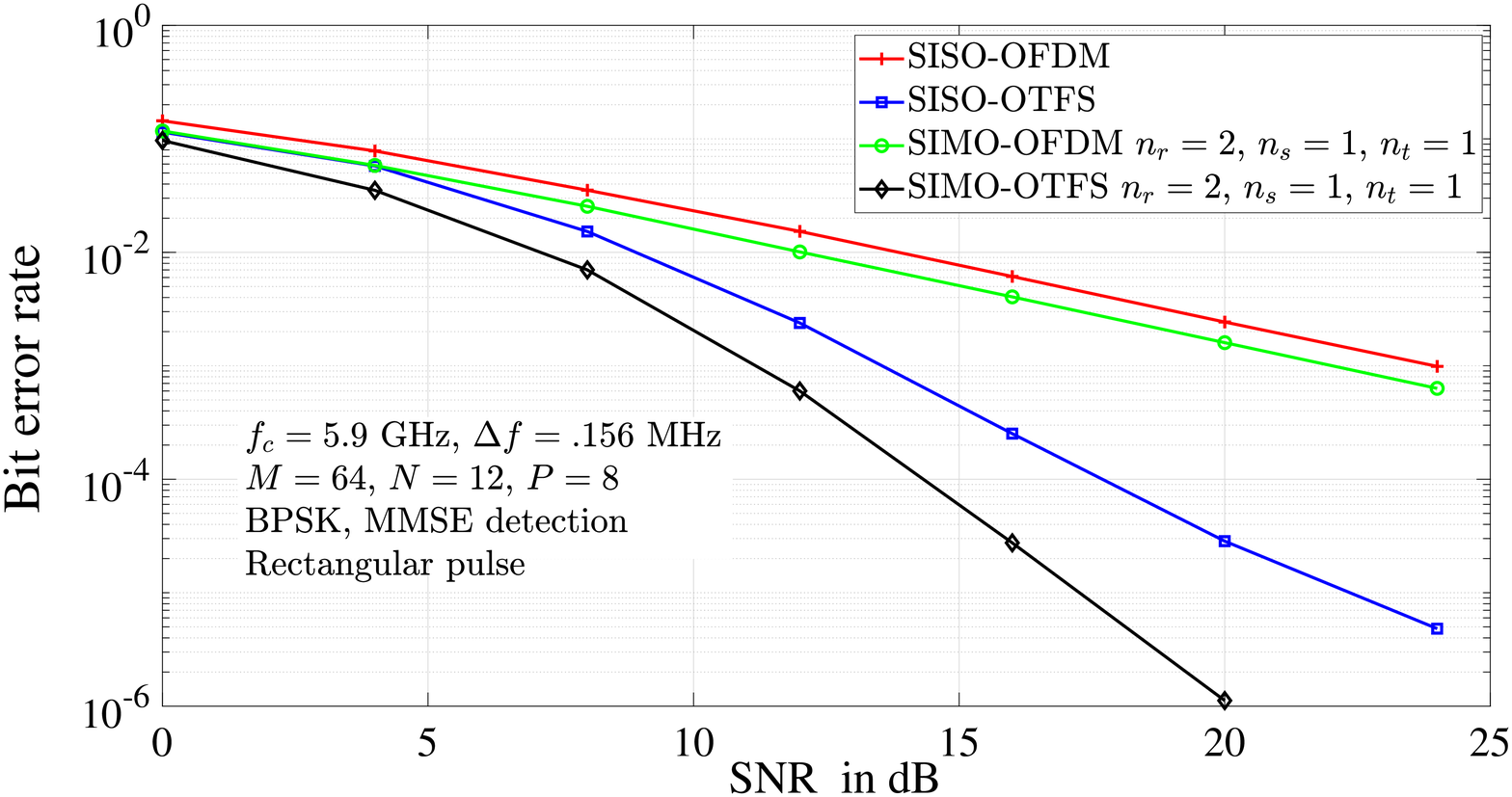}
\vspace{-2mm}
\caption{BER performance comparison between SIMO-OTFS with RAS and SIMO-OFDM with RAS for $M=64$, $N=12$, $P=8$, $n_s=1$, $n_r=1,2$, MMSE detection, and fractional delays/Dopplers.}
\label{OTFS_OFDM_mmse}
\vspace{-4mm}
\end{figure}

{\em Performance in IEEE 802.11p with rectangular pulse:} Here, we present a performance comparison between SIMO-OTFS and SIMO-OFDM with RAS considering system parameters according to IEEE 802.11p standard \cite{WAVE} as follows. The carrier frequency and subcarrier spacing are taken to be 5.9 GHz and 0.156 MHz, respectively. A frame size of $M=64$, $N=12$, number of paths $P=8$, and a maximum speed of 220 km/h (corresponding maximum Doppler of 1.2 kHz), and BPSK modulation are considered. Figure \ref{OTFS_OFDM_mmse} shows the performance comparison between SIMO-OTFS with rectangular pulse and SIMO-OFDM for $M=64$, $N=12$, $P=8$, $n_s=1$, $n_r=2$, and MMSE detection. From Fig. \ref{OTFS_OFDM_mmse}, we observe that the performance of SIMO-OTFS with RAS is significantly better than that of SIMO-OFDM with RAS. For example, at a BER of $10^{-3}$, SIMO-OTFS with RAS has an SNR gain of about 11 dB compared to SIMO-OFDM with RAS.

\begin{figure}
\includegraphics[width=9cm,height=6cm]{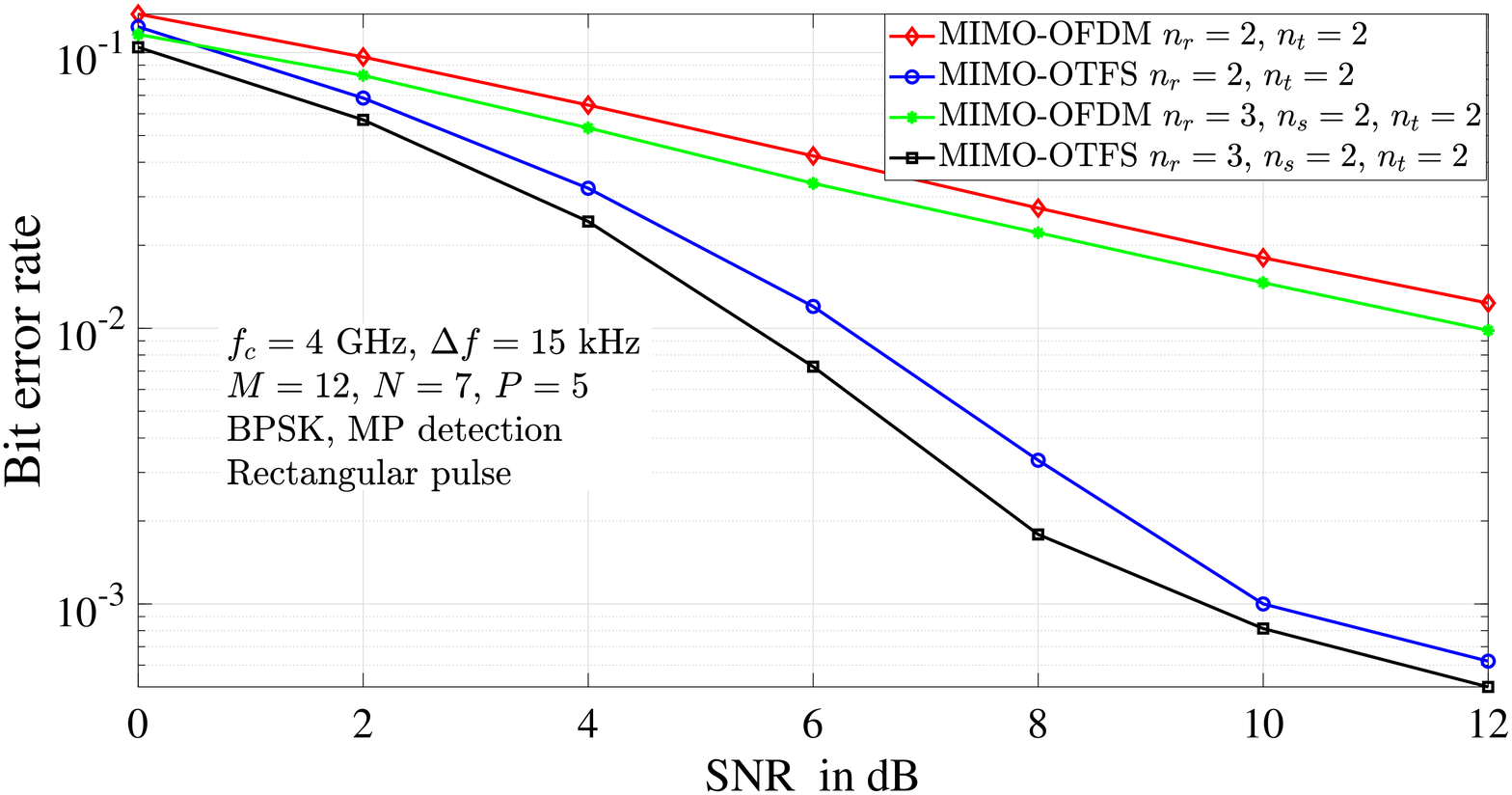}
\vspace{-2mm}
\caption{BER performance comparison between MIMO-OTFS with RAS and MIMO-OFDM with RAS for $M=12$, $N=7$, $P=5$, $n_t=2$, $n_s=2$, $n_r=2,3$, MP detection, and fractional delays/Dopplers.}
\vspace{-4mm}
\label{OTFS_OFDM_mp}
\end{figure}

{\em Performance in LTE with rectangular pulse:} 
Here, we present a performance comparison between MIMO-OTFS and MIMO-OFDM with RAS considering system parameters according to LTE standard \cite{lte} as follows. The carrier frequency and subcarrier spacing are taken to be 4 GHz and 15 kHz, respectively. A frame size of $M=12$, $N=7$, $P=5$, and a maximum speed of 500 km/h (corresponding maximum Doppler of 1.85 kHz), and BPSK modulation are considered. Figure \ref{OTFS_OFDM_mp} shows the performance comparison between MIMO-OTFS with rectangular pulse and MIMO-OFDM for $M=12$, $N=7$, $P=5$, $n_t=2$, $n_s=2$, $n_r=2,3$, and MP detection. From Fig. \ref{OTFS_OFDM_mp}, we observe that MIMO-OTFS with RAS performs better than MIMO-OFDM with RAS. We further note that while the performance for $P>1$ in Figs. \ref{OTFS_OFDM_mmse} and \ref{OTFS_OFDM_mp} are observed through simulations, an analytical derivation of the diversity orders for $P>1$ with RAS for the fractional delay-Doppler case is open for future investigation.

\end{document}